\documentclass[journal]{IEEEtran}

\ifCLASSINFOpdf
\else
   \usepackage[dvips]{graphicx}
\fi
\usepackage{url}

\usepackage{etoolbox}%
\robustify\bfseries
\usepackage{graphicx}
\usepackage{amsmath,amssymb,graphicx}
\usepackage[]{algorithm2e}
\usepackage{tabularx}
\usepackage{wasysym}
\usepackage{color,soul}
\usepackage{cite}
\usepackage{balance}
\usepackage{booktabs}
\usepackage{multirow}
\usepackage{siunitx}
\usepackage{breqn}
\usepackage{xcolor}
\usepackage{makecell}
\def\H{{\mathsf H}}
\def\T{{\mathsf T}}
\def\CC{{\mathbb C}}

\usepackage{makecell}
\usepackage{caption}
\usepackage{arydshln}
\usepackage{fixltx2e}
\usepackage{hyperref}

\begin{document}

\title{Leveraging Low-Distortion Target Estimates for Improved Speech Enhancement}

\author{
Zhong-Qiu Wang, Gordon Wichern, and Jonathan Le Roux
\thanks{Manuscript received on Oct. 1, 2021.}
\thanks{
Z.-Q. Wang was with Mitsubishi Electric Research Laboratories (MERL), Cambridge, MA 02139, USA, while performing this work.
He is now with the Language Technologies Institute, Carnegie Mellon University, Pittsburgh, PA 15213, USA (e-mail: wang.zhongqiu41@gmail.com).}
\thanks{
G.\ Wichern and J.\ Le Roux are with MERL, Cambridge, MA 02139, USA (e-mail: \{wichern,leroux\}@merl.com).}
}

\markboth{}
{Shell \MakeLowercase{\textit{et al.}}: Bare Demo of IEEEtran.cls for IEEE Journals}
\maketitle

\begin{abstract}

A promising approach for multi-microphone speech separation involves two deep neural networks (DNN), where the predicted target speech from the first DNN is used to compute signal statistics for time-invariant minimum variance distortionless response (MVDR) beamforming, and the MVDR result is then used as extra features for the second DNN to predict target speech.
Previous studies suggested that the MVDR result can provide complementary information for the second DNN to better predict target speech.
However, on fixed-geometry arrays, both DNNs can take in, for example, the real and imaginary (RI) components of the multi-channel mixture as features to leverage the spatial and spectral information for enhancement.
It is not explained clearly why the linear MVDR result can be complementary and why it is still needed, considering that the DNNs and the beamformer use the same input, and the DNNs perform non-linear filtering and could render the linear filtering of MVDR unnecessary.
Similarly, in monaural cases, one can replace the MVDR beamformer with a monaural weighted prediction error (WPE) filter.
Although the linear WPE filter and the DNNs use the same mixture RI components as input, the WPE result is found to significantly improve the second DNN.
This study provides a novel explanation from the perspective of the low-distortion nature of such algorithms, and finds that they can consistently improve phase estimation.
Equipped with this understanding, we investigate several low-distortion target estimation algorithms including several beamformers, WPE, forward convolutive prediction (FCP), and their combinations, and use their results as extra features to train the second network to achieve better enhancement.
Evaluation results on single- and multi-microphone speech dereverberation and enhancement tasks indicate the effectiveness of the proposed approach, and the validity of the proposed view.

\end{abstract}


\IEEEpeerreviewmaketitle

\section{Introduction}

\IEEEPARstart{R}{oom} reverberation and environmental noise are pervasive in modern hands-free speech communication applications such as teleconferencing, hearing aids, smart speakers, and robust automatic speech recognition (ASR).
They can dramatically degrade speech intelligibility and quality, and are very detrimental to modern ASR systems.
Speech enhancement using a single or an array of microphones is desirable for such applications.
In the past decade, deep learning based approaches have been firmly established as the state-of-the-art approach for speech enhancement \cite{WDLreview}.
Given a single-speaker utterance recorded in a noisy-reverberant environment by a $P$-microphone array, the physical model in the short-time Fourier transform (STFT) domain can be formulated as
\begin{align} 
	\mathbf{Y}(t,f) &= \mathbf{X}(t,f)+\mathbf{N}(t,f) \nonumber \\
	&= \mathbf{S}(t,f)+\mathbf{H}(t,f)+\mathbf{N}(t,f) \nonumber \\
	&= \mathbf{S}(t,f)+\mathbf{V}(t,f), \label{eq:phymodel_freq}
\end{align}
where $\mathbf{Y}(t,f)$, $\mathbf{N}(t,f)$, $\mathbf{X}(t,f)$, $\mathbf{S}(t,f)$ and $\mathbf{H}(t,f)\in \CC^{P}$ respectively denote the STFT vectors of the mixture, reverberant noise, reverberant speech, direct and non-direct signals of the target speaker, at time $t$ and frequency $f$.
A speech enhancement system usually aims at recovering the target speaker's direct-path signal $S_q$ captured at a reference microphone $q$ while reducing the other signals $\mathbf{V}=\mathbf{H}+\mathbf{N}$, based on the multi-channel input $\mathbf{Y}$.
Note that variables without $t$ and $f$ refer to the corresponding spectrogram.

\begin{figure}
  \centering  
  \includegraphics[width=9cm]{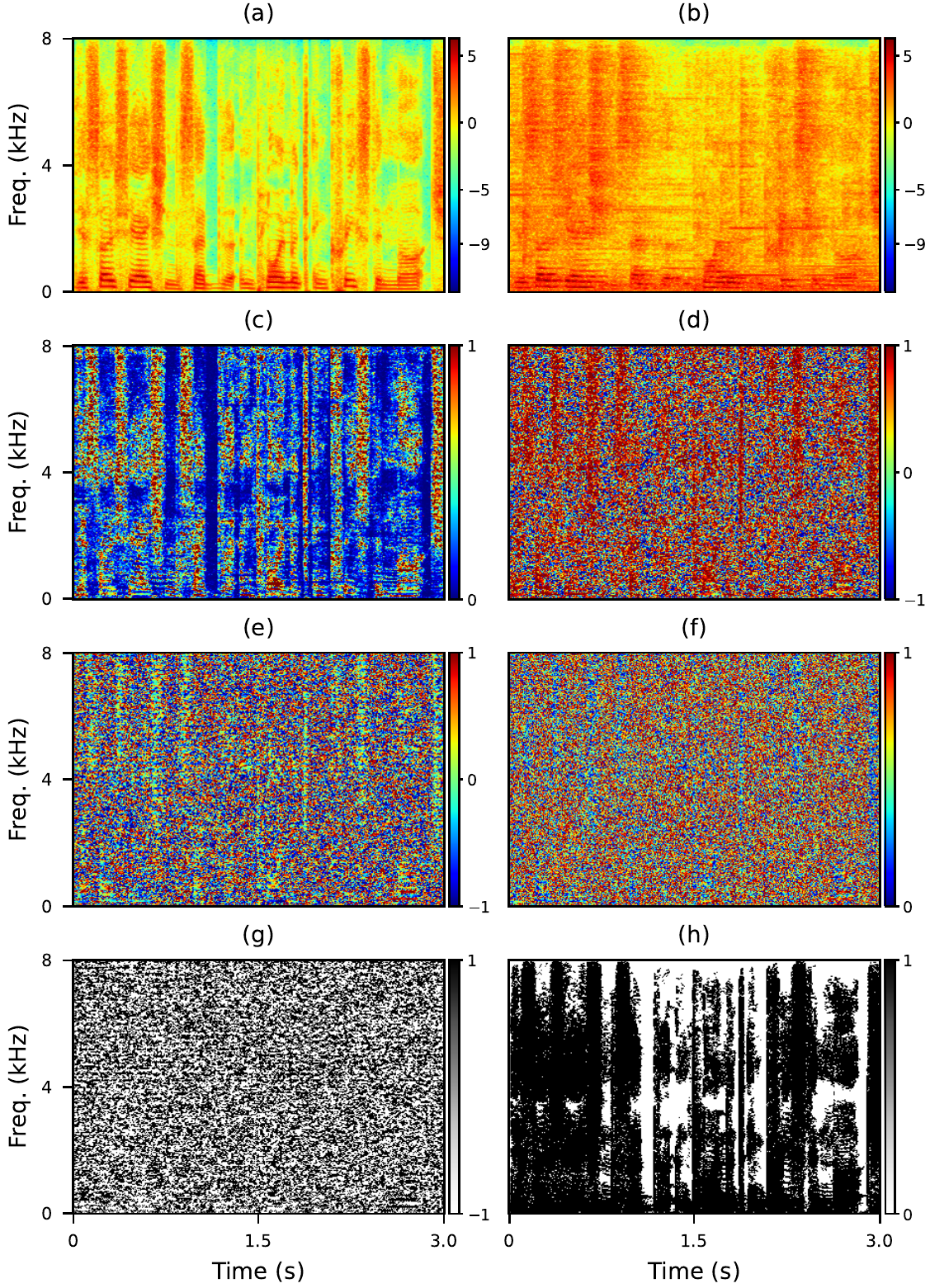}
  \vspace{-0.6cm}
  \caption{\footnotesize{
  Illustration of an example
  (a) target power spectrogram $\text{log}(|S_q|)$;
  (b) mixture power spectrogram $\text{log}(|Y_q|)$;
  (c) spectral magnitude mask $|S_q|/|Y_q|$ truncated to be below one;
  (d) $\cos(\angle S_q - \angle Y_q)$;
  (e) $\sin(\angle S_q - \angle Y_q)$;
  (f) absolute of $\sin(\angle S_q - \angle Y_q)$;
  (g) sign of $\angle e^{j(\angle S_q - \angle Y_q)}$;
  (h) binary mask denoting T-F units with active target speech.
  Best viewed in color.
  }}
  \vspace{-0.45cm}
  \label{patternsfigure}
\end{figure}

To estimate $S_q$, a popular approach trains a DNN to estimate the ideal complex ratio mask \cite{Williamson2016, WDLreview}, which can perfectly reconstruct the target speech.
It is defined as
\begin{align} 
	M_q &= S_q/Y_q = |S_q|/|Y_q|\cos( \theta_q ) + j|S_q|/|Y_q|\sin( \theta_q ),
\end{align}
where $\theta_q = \angle S_q - \angle Y_q$ is the phase difference between the target speech and the mixture, and $j$ the imaginary unit.
The real component of $M_q$, also known as the non-truncated phase-sensitive mask \cite{Erdogan2015}, is the product of the magnitude ratio $|S_q|/|Y_q|$ and the cosine phase difference $\cos\big( \theta_q \big)$.
$|S_q|/|Y_q|$, known as the spectral magnitude mask \cite{WYXtrainingtargets} (or ideal amplitude mask \cite{Erdogan2015}), can be reasonably predicted based on the mixture magnitude \cite{WDLreview}, since both $|S_q|$ and $|Y_q|$ exhibit strong spectro-temporal patterns that can be learned by a supervised learning based model. 
See Fig.~\ref{patternsfigure}(a), (b), and (c) for an illustration of the patterns of an example noisy-reverberant mixture.
As shown in Fig.~\ref{patternsfigure}(d), $\cos\big( \theta_q \big)$ exhibits some patterns similar to those in the spectral magnitude mask.
This is because, as the input signal-to-noise ratio (SNR) becomes lower, $\angle Y_q(t,f)$ gets closer to $\angle V_q(t,f)$, which is likely different from $\angle S_q(t,f)$, and hence $\cos\big( \theta_q(t,f) \big)$ likely becomes smaller than one.
However, knowing exactly how much smaller requires estimating the absolute value of $\theta_q(t,f)$, and is a difficult task.
The imaginary component of $M_q$ also includes $|S_q|/|Y_q|$, which can be reasonably predicted.
However, $\sin(\theta_q)$, illustrated in Fig.~\ref{patternsfigure}(e), appears difficult to predict, due to the lack of clear patterns.
Indeed, from a predicted $\cos(\hat{\theta}_q)$, one can compute $|\sin(\hat\theta_q)|$ (see Fig.~\ref{patternsfigure}(f)) as $\sqrt{1-\cos(\hat\theta_q)^2}$, but this is only the absolute value.
To accurately estimate $\sin(\theta_q)$, one also has to  estimate the sign of the phase difference between $S_q$ and $Y_q$ at each T-F unit. 
This is however known to be a difficult task \cite{WZQtrigonometric2019}.
Fig.~\ref{patternsfigure}(g) plots the sign of the phase difference.
Clearly, the pattern is very random, simply because the sign of $\theta_q(t,f)$ depends on the phase of the non-target signal $V_q(t,f)$.
To accurately estimate the target phase, a successful algorithm should be capable of estimating the phase difference at each T-F unit in terms of its sign and absolute value, either explicitly or implicitly.

\begin{figure}
  \centering  
  \includegraphics[width=4.25cm]{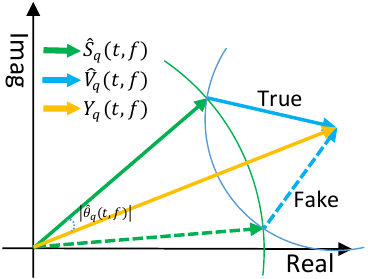}
  \vspace{-0.1cm}
  \caption{\footnotesize{
  Illustration of phase-difference sign ambiguity in the complex plane.}}
  \label{ambiguityfigure}
  \vspace{-0.4cm}
\end{figure}

Our preliminary study \cite{WZQtrigonometric2019} proposed a DNN-based algorithm to explicitly predict the sign, and implicitly predict the absolute value of the phase difference through magnitude estimation.
The key idea is that if the magnitude of $S_q$ and $V_q$ can be accurately estimated (in the oracle case: let's assume $|\hat{S}_q|=|S_q|$ and $|\hat{V}_q|=|V_q|$) and if $\hat{S}_q$ and $\hat{V}_q$ add up to the mixture (i.e., $Y_q=\hat{S}_q+\hat{V}_q$), the absolute phase difference can be uniquely determined based on the law of cosines (see Fig.~\ref{ambiguityfigure}), and the phase solution at each T-F unit can be narrowed down to only two candidates:
\begin{align} 
	|\hat\theta_q(t,f)| & =  \arccos \Big(\frac{|Y_q(t,f)|^2+|\hat{S}_q(t,f)|^2-|\hat{V}_q(t,f)|^2}{2|Y_q(t,f)||\hat{S}_q(t,f)|}\Big) \\
	\angle \hat{S}_q(t,f) &= \angle Y_q(t,f) \pm |\theta_q(t,f)|.
\end{align}
Based on this insight, the DNNs in \cite{WZQtrigonometric2019} are designed to predict the magnitudes of target and non-target signals and the phase-difference sign, and at the same time to enforce the predicted target and non-target signals to sum up to the mixture.
However, the sign is found to be very difficult to predict accurately.
This is partly because at each T-F unit, we only observe the mixture vector $Y(t,f)$, while there are two possible target and non-target pairs producing the same mixture but being symmetric with respect to the mixture vector (see Fig.~\ref{ambiguityfigure}).
Intuitively, this ambiguity occurs because the phase of the target source could be either ahead of or behind the mixture phase at each T-F unit in an almost random way (see Fig.~\ref{patternsfigure}(g)).
This randomness could pose fundamental difficulties for supervised learning based phase estimation \cite{WZQtrigonometric2019}, as supervised learning based models usually require clear spectro-temporal patterns in order to learn to make predictions.
The root cause of this randomness is likely because, in monaural cases, we only observe one signal (i.e., the mixture), but we want to reconstruct multiple signals (i.e., the sources).
This is an ill-posed problem in nature.

\begin{figure}
  \centering  
  \includegraphics[width=8.5cm]{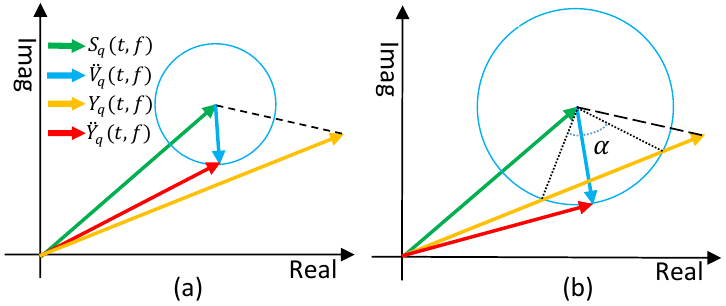}
  \vspace{-0.1cm}
  \caption{\footnotesize{
  Complex-plane illustration of benefits of low-distortion target estimates when non-target signals are (a) sufficiently suppressed; (b) not sufficiently suppressed. Best viewed in color.}}
  \label{addressingambiguityfigure}
  \vspace{-0.4cm}
\end{figure}

In this context, we propose to leverage low-distortion target estimates produced by conventional single- or multi-microphone enhancement algorithms to improve deep learning based phase estimation.
Our insight is that if we can first apply a distortionless enhancement algorithm to the mixture such that the target signal is maintained distortionlessly while the non-target signal is suppressed to some extent, the processed mixture $\ddot{Y}_q$ (or $\hat{S}_q$) would indicate the phase-difference sign.
To illustrate this idea, let us denote the processed mixture as 
\begin{align} 
	\ddot{Y}_q(t,f) = \hat{S}_q(t,f) = S_q(t,f) + \ddot{V}_q(t,f),
\end{align}
where the target speech $S_q$ is distortionlessly maintained, and $\ddot{V}_q$ denotes the suppressed non-target signal with $|\ddot{V}_q(t,f)| < |V_q(t,f)|$ for many T-F units.
Fig.~\ref{addressingambiguityfigure}(a) illustrates the case when $|\ddot{V}_q(t,f)|$ is much smaller than $|V_q(t,f)|$.
Notice that $\ddot{V}_q(t,f)$ could have any phase value, so the blue vector could point in any direction.
In this case, regardless of the phase value of $\ddot{V}_q(t,f)$, $\ddot{Y}_q(t,f)$ always indicates that the true target phase advances the mixture phase and hence the sign ambiguity can be resolved.
Fig.~\ref{addressingambiguityfigure}(b) illustrates the case when $|\ddot{V}_q(t,f)|$ is not much smaller than $|V_q(t,f)|$.
In this case, $\angle \ddot{Y}_q(t,f)$ could be on the wrong side.
However, this may only happen when $ |\ddot{V}_q(t,f)| > |S_q(t,f)| \sin( \theta_q(t,f)) $, with a likely-small probability $\frac{\alpha}{2\pi}$ (see the figure for the definition of $\alpha$) which can be computed via simple trigonometry as $\frac{1}{\pi}\arccos \Big(\frac{|S_q(t,f)|}{|\ddot{V}_q(t,f)|}\sin( \theta_q(t,f))\Big)$.
This probability is small in cases where $|\ddot{V}_q(t,f)|$ is well suppressed by the low-distortion algorithm, or where $\theta_q(t,f)$ is large, which is the case where getting the sign right matters most.
Besides the benefits of resolving sign ambiguity, $\ddot{Y}_q(t,f)$ is expected to be closer than $Y_q(t,f)$ to $S_q(t,f)$ at many T-F units, simply because the target speech is distortionlessly maintained while non-target signals are suppressed.
Therefore, it could also be very helpful at determining the absolute value of the phase difference and improving the estimation of the target magnitude.

Equipped with this novel understanding, we propose a 2stage-DNN approach for speech enhancement.
The first network is trained to predict the target speech.
The predicted speech is then used to compute signal statistics for beamforming and WPE- \cite{Yoshioka2012} or FCP-based \cite{Wang2021FCPjournal} dereverberation.
All of them are linear, time-invariant, and known to produce low-distortion target estimates \cite{Yoshioka2012, WDLreview, Haeb-Umbach2020, Wang2021FCPjournal}.
We then use their estimates as extra features to train the second network to better predict target speech.
Different from our earlier work \cite{WZQtrigonometric2019}, where DNNs are trained to explicitly predict the target magnitude, absolute phase difference, and phase-difference sign, this work trains DNNs in the complex T-F domain to predict the RI components of the target speech, hence implicitly predicting the target phase.
This could better leverage the power of deep learning based end-to-end optimization.

Our study makes two major contributions.
First and foremost, we provide a novel view that low-distortion target estimates produced by conventional enhancement algorithms could be very helpful at improving phase estimation.
This view provides a good understanding on the reason why our approach works, and reveals its strong potential.
Second, we explore and compare a number of ways to obtain low-distortion target estimates, including beamforming, and WPE- or FCP-based dereverberation.
Our systems build upon a strong 2stage-DNN MISO-BF-MISO baseline \cite{Wang2020c, Wang2020d}, which needs to run the first network once for each microphone, and requires a uniform circular array geometry.
The proposed systems only run the first network once for a reference microphone to reduce the amount of computation, and avoid the reliance on that particular type of array geometry.
We also provide some other minor contributions, which will be stated when discussing specific techniques.
We shall note that part of this work has been published %
in ICASSP 2020 \cite{Wang2020d}, which only deals with speech dereverberation and only considers using MVDR beamformers to obtain low-distortion target estimates.

The rest of this paper is organized as follows.
We provide a system overview in Section~\ref{systemoverview}, followed by a baseline system in Section~\ref{baseline},  our proposed systems in Section~\ref{proposedsystems}, and DNN configurations in Section~\ref{dnndescription}.
We present experimental setup and evaluation results in Sections~\ref{setup} and \ref{results}, and draw conclusions in Section~\ref{conclusion}.

\section{System Overview}\label{systemoverview}

\begin{figure}
  \centering
  \includegraphics[width=8.5cm]{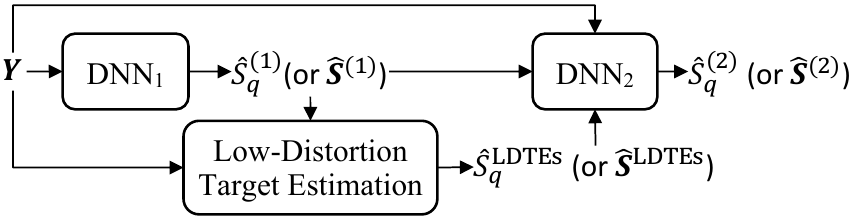}
  \caption{System overview.}
  \label{overviewfigure}
  \vspace{-0.6cm}
\end{figure}

Figure~\ref{overviewfigure} illustrates the high-level architecture of our system.
Different variations of this system will be presented with more details later in Sections~\ref{baseline} and \ref{proposedsystems}.
It contains two DNNs and in between a module that can produce low-distortion target estimates.
Based on the multi-channel mixture $\mathbf{Y}$, the first DNN estimates target anechoic speech at all the microphones (i.e., $\hat{\mathbf{S}}^{(1)}$) or just at the reference microphone $q$ (i.e., $\hat{S}_q^{(1)}$).
We then combine the mixture and the outputs from the first DNN and the low-distortion target estimation (LDTE) module as inputs for a second DNN to further estimate target anechoic speech.
Both DNNs are trained using single- or multi-microphone complex spectral mapping \cite{Wang2020b, Wang2020a, Wang2020c}, where we predict the RI components of target speech from the mixture RI components.
More DNN details will be provided in Section \ref{dnndescription}.
At this point, readers can assume that each DNN in our system can provide an estimate of target anechoic speech in the complex T-F domain, denoted as $\hat{\mathbf{S}}^{(b)}$ or $\hat{S}_q^{(b)}$, where $b\in\{1,2\}$ as there are two DNNs.

There are many options for the LDTE module.
This study considers DNN-supported WPE \cite{Kinoshita2017}, FCP \cite{Wang2021FCPjournal}, and beamformers \cite{Heymann2015, Yoshioka2015, Heymann2016a, Zhang2017a} as well as their combinations, where DNN-provided signal statistics are used for filter estimation.

Following \cite{Wang2020d, Wang2020c}, we assume the same array geometry is used for training and testing.
This is a valid assumption, since in products such as Amazon Echo and Google Home, the number of microphones and their arrangement are fixed.
In addition, we assume offline processing scenarios.

\section{Baseline: MISO$_1$+MVDR+MISO$_2$ System}\label{baseline}

\begin{figure}
  \centering  
  \includegraphics[width=8.5cm]{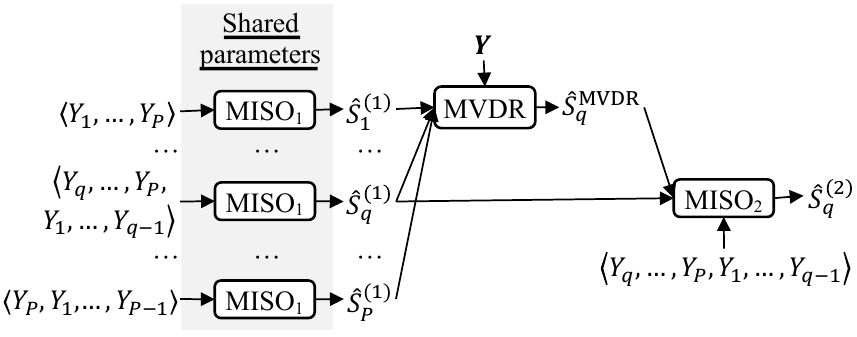}
  \caption{MISO$_1$+MVDR+MISO$_2$ system.}
  \label{MISO1+MVDR+MISO2}
  \vspace{-0.3cm}
\end{figure}

Figure~\ref{MISO1+MVDR+MISO2} illustrates %
the MISO-BF-MISO system proposed in \cite{Wang2020c, Wang2020d}.
MISO denotes a \textit{multi-microphone input and single-microphone output} network, where the multi-channel mixture is included as input to the network to predict the target speech at the reference microphone.
More specifically, the MISO$_1$ network is trained to predict the RI components of $S_q$ based on the RI components of an ordered concatenation of $\big[Y_q, \dots, Y_P, Y_1, \dots, Y_{q-1}\big]$.
Assuming a uniform circular geometry, at run time we can circularly shift the multi-microphone input to predict the target speech captured at each microphone.
For example, we can feed $\big[Y_1, \dots, Y_P\big]$ to MISO$_1$ to obtain $\hat{S}_1^{(1)}$, feed $\big[Y_2, \dots, Y_P, Y_1\big]$ to obtain $\hat{S}_2^{(1)}$, and so on.
After obtaining $\hat{\mathbf{S}}^{(1)}(t,f)=\big[\hat{S}_1^{(1)}(t,f), \dots, \hat{S}_P^{(1)}(t,f)\big]^{\T}$, we use it to derive an MVDR beamformer.
The beamforming result $\hat{S}_q^{\text{MVDR}}$ is then combined with $\hat{S}_q^{(1)}$ and the mixture $\big[Y_q, \dots, Y_P, Y_1, \dots, Y_{q-1}\big]$ as the input to MISO$_2$, which can be considered as a post-filtering network, to estimate the target speech again.
Note that we use different subscripts in, say, MISO$_1$ and MISO$_2$ to differentiate different models we trained.
This convention applies to all the models in this study.

\begin{figure}
  \centering  
  \includegraphics[width=7cm]{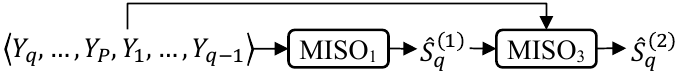}
  \caption{MISO$_1$+MISO$_3$ system.}
  \label{MISO1+MISO3}
  \vspace{-0.3cm}
\end{figure}

The MVDR beamformer is computed as follows.
Given $\hat{\mathbf{S}}^{(1)}$, the target and non-target covariance matrices, $\hat{\mathbf{\Phi}}^{(s)}(f)$ and $\hat{\mathbf{\Phi}}^{(v)}(f)$, are computed as
\begin{align}
\hat{\mathbf{\Phi}}^{(s)}(f) &= \sum\nolimits_t \hat{\mathbf{S}}^{(1)}(t,f)\hat{\mathbf{S}}^{(1)}(t,f)^{\H}, \label{covariancematrixS} \\
\hat{\mathbf{\Phi}}^{(v)}(f) &= \sum\nolimits_t \hat{\mathbf{V}}^{(1)}(t,f)\hat{\mathbf{V}}^{(1)}(t,f)^{\H}, \label{covariancematrixV} \\
\hat{\mathbf{V}}^{(1)}(t,f) &= \mathbf{Y}(t,f)-\hat{\mathbf{S}}^{(1)}(t,f).
\end{align}
Following \cite{Yoshioka2015, Zhang2017a}, the steering vector $\hat{\mathbf{d}}(f)$ of the target speaker is computed as
\begin{align}\label{steeringvec}
\hat{\mathbf{d}}(f)=\mathcal{P}\big(\hat{\mathbf{\Phi}}^{(s)}(f)\big),
\end{align}
where $\mathcal{P}(\cdot)$ extracts the principal eigenvector.
Designating microphone $q$ as the reference, an MVDR beamformer is computed as
\begin{align}\label{mvdrfilter}
\hat{\mathbf{w}}(f;q) = \frac{\hat{\mathbf{\Phi}}^{(v)}(f)^{-1} \hat{\mathbf{d}}(f)}{\hat{\mathbf{d}}(f)^{\H} \hat{\mathbf{\Phi}}^{(v)}(f)^{-1} \hat{\mathbf{d}}(f)} \hat{d}_q^{*}(f),
\end{align}
where $(\cdot)^{*}$ computes complex conjugate, and the beamforming result is computed as
\begin{align}\label{mvdrresult}
\hat{S}_q^{\text{MVDR}}(t,f) = \hat{\mathbf{w}}(f;q)^{\H}\mathbf{Y}(t,f).
\end{align}

Note that if the beamformer is computed using oracle statistics, the oracle MVDR (oMVDR) result is
\begin{align}\label{mvdrresultoracle}
&\hat{S}_q^{\text{oMVDR}}(t,f) = \mathbf{w}(f;q)^{\H}\mathbf{Y}(t,f) \nonumber \\
&= \!\Big(\frac{\mathbf{\Phi}^{(v)}(f)^{-1} \mathbf{d}(f)}{\mathbf{d}(f)^{\H} \mathbf{\Phi}^{(v)}(f)^{-1} \mathbf{d}(f)} d_q^{*}(f)\!\Big)^{\H} \Big(\frac{\mathbf{d}(f)}{d_q(f)}S_q(t,f)\!+\!\mathbf{V}(t,f)\!\Big) \nonumber \\
&= S_q(t,f) + \Big(\frac{\mathbf{\Phi}^{(v)}(f)^{-1} \mathbf{d}(f)}{\mathbf{d}(f)^{\H} \mathbf{\Phi}^{(v)}(f)^{-1} \mathbf{d}(f)} d_q^{*}(f)\Big)^{\H} \mathbf{V}(t,f).
\end{align}
As we can see, ideally the target signal is maintained distortionless while non-target signals are suppressed.
The estimated result obtained in Eq.~(\ref{mvdrresult}) is expected to have low distortion, as long as the estimated statistics are reasonably good.

This 2stage-DNN approach with an MVDR module in between has shown strong performance in tasks such as speech enhancement \cite{Wang2020a}, speech dereverberation \cite{Wang2020b, Wang2020d}, and speaker separation \cite{Wang2020c}.
It has shown better performance than a MISO$_1$+MISO$_3$ system that simply stacks two MISO networks (see Fig.~\ref{MISO1+MISO3}).
The inclusion of an MVDR beamformer is usually perceived as an ensemble approach, where the second DNN can integrate spatial and spectral features for separation.
A critique of this approach is that the beamformer and the MISO networks use the same input (i.e., $\mathbf{Y}$), and the MVDR beamformer is just a simple linear filter and therefore could be unnecessary, especially when the array geometry is fixed, as a MISO network can be viewed as a powerful non-linear beamformer, which could potentially replace conventional linear beamformers.
Although earlier studies \cite{Wang2020c, Wang2020d} experimentally show that using an MVDR beamformer in between the two networks leads to consistent improvement, and suggest that the MVDR result can provide complementary information to plain DNN-based end-to-end modeling, what information is complementary and why it is complementary is not analyzed, and there lacks a fundamental understanding on why the MVDR beamformer can lead to consistent improvement.
Our analysis in the introduction provides an explanation, suggesting that a beamformer would likely be helpful, as its low-distortion estimates could help a DNN to better predict the target speech, especially its phase.

\section{Proposed Systems}\label{proposedsystems}

MVDR is one way of obtaining low-distortion target estimates.
We propose to leverage other low-distortion algorithms to compute extra features to train the second network, as better low-distortion target estimates are likely to improve the second DNN.
This section investigates various
beamformers, and their integration with the DNN-WPE and DNN-FCP algorithms.
In addition, we propose mechanisms that can avoid running the first DNN multiple times to reduce the amount of computation, and avoid the reliance on uniform circular array geometry.

\subsection{MISO$_1$+MVDR+WPE+MISO$_4$ System}

DNN-WPE \cite{Nakatani2010, Kinoshita2017} computes a filter to linearly combine past observations to estimate the late reverberation at the current frame, and the dereverberation result is obtained by subtracting the estimate from the mixture, i.e.,
\begin{align}\label{wperesult}
\hat{S}_q^{\text{WPE}}(t,f)=Y_q(t,f)-\hat{\mathbf{g}}(f;q)^{\H}\widetilde{\mathbf{Y}}(t-\Delta,f),
\end{align}
where $\widetilde{\mathbf{Y}}(t,f)=[\mathbf{Y}(t,f)^\T,\dots,\mathbf{Y}(t-K+1,f)^\T]^\T$, $K$ is the filter taps, $\hat{\mathbf{g}}(f;q)\in \CC^{KP}$ a $KP$-dimensional filter, and $\Delta$ ($\geq 1$) a prediction delay.
Eq.~(\ref{wperesult}) is equivalent to
\begin{align}\label{wperesultanalysis}
\hat{S}_q^{\text{WPE}}(t,f)=S_q(t,f)+\Big(V_q(t,f)-\hat{\mathbf{g}}(f;q)^{\H}\widetilde{\mathbf{Y}}(t-\Delta,f)\Big),
\end{align}
where, because of the non-zero $\Delta$, $\hat{\mathbf{g}}(f;q)^{\H}\widetilde{\mathbf{Y}}(t-\Delta,f)$ would likely only approximate late reverberation contained in $V_q(t,f)$, and avoid cancelling target speech \cite{Nakatani2010}.
Therefore, the target signal is expected to be distortionlessly maintained while non-target signals are suppressed.

\begin{figure}
  \centering  
  \includegraphics[width=8.5cm]{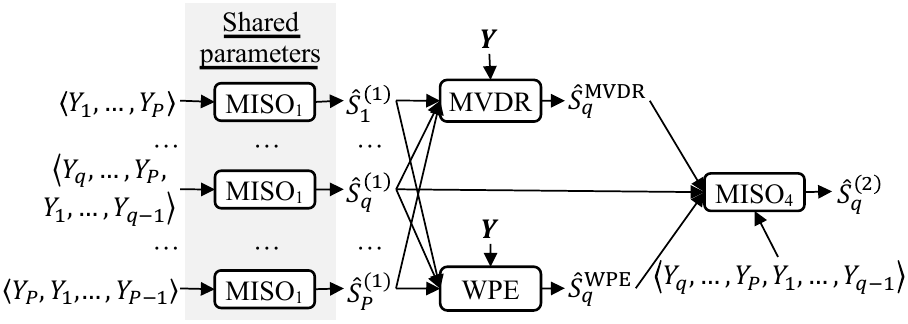}
  \caption{MISO$_1$+MVDR+WPE+MISO$_4$ system.}
  \label{MISO1+MVDR+WPE+MISO4}  \vspace{-0.4cm}
\end{figure}

Our study includes a DNN-WPE \cite{Kinoshita2017} result to train the second network (see Fig.~\ref{MISO1+MVDR+WPE+MISO4}).
The filter is computed by optimizing a quadratic objective as follows \cite{Kinoshita2017}:
\begin{align}\label{dnnwpe}
\underset{\mathbf{g}(f;q)}{{\text{argmin}}} \sum\nolimits_{t} \frac{|Y_q(t,f)-\mathbf{g}(f;q)^{\H}\ \widetilde{\mathbf{Y}}(t-\Delta,f)|^2}{\hat{\lambda}(t,f)}.
\end{align}
Following \cite{Kinoshita2017}, we compute the power spectral density (PSD) $\hat{\lambda}$ based on DNN outputs as
\begin{align}\label{lambdaallestimates}
\hat{\lambda}(t,f) =\text{max}(\varepsilon \text{max}(\sum\nolimits_q |\hat{S}_q^{(1)}|^2),\sum\nolimits_q |\hat{S}_q^{(1)}(t,f)|^2),
\end{align}
where $\text{max}(\cdot)$ extracts the maximum value of a spectrogram, $\text{max}(\cdot,\cdot)$ returns the larger of two values, and $\varepsilon$ is a floor value to avoid placing too much weight on T-F units without significant target energy.

Our contribution here is including DNN-WPE results to train the second network.
Both MVDR and DNN-WPE results are low-distortion.
Their parallel combination could help determine the phase-difference sign and predict target phase and magnitude.
We will discuss a cascaded combination of WPE and beamforming in Section~\ref{cascadeWPEandBF}.

\subsection{MIMO+MVDR+WPE+MISO$_5$ System}

\begin{figure}
  \centering  
  \includegraphics[width=8.5cm]{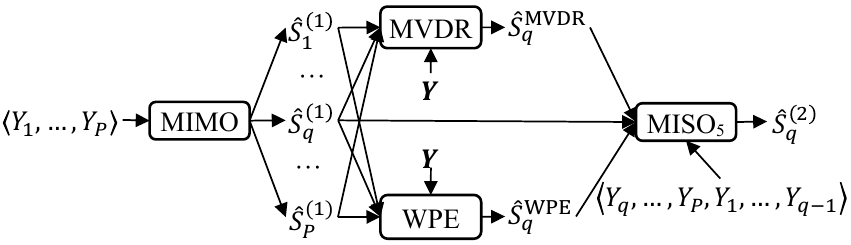}
  \caption{MIMO+MVDR+WPE+MISO$_5$ system.}
  \label{MIMO+MVDR+WPE+MISO5}\vspace{-0.4cm}
\end{figure}

In Figs.~\ref{MISO1+MVDR+MISO2} and \ref{MISO1+MVDR+WPE+MISO4}, at run time the MISO$_1$ network is run once for each microphone to get the statistics for beamforming and WPE.
The amount of computation is high and possibly unnecessary.
To reduce it, we propose to use a \textit{multi-microphone input and multi-microphone output} (MIMO) network to directly predict the target speech at all the microphones.
See Fig.~\ref{MIMO+MVDR+WPE+MISO5} for an illustration.
Compared with MISO$_1$+MVDR+WPE+MISO$_4$, MIMO+MVDR+WPE+MISO$_5$ does not require microphones to be arranged in a uniform circular way.
In our experiments (and also in our preliminary study \cite{Wang2020d}), at each microphone the predicted speech by MIMO is found to be worse than MISO$_1$, likely because in MIMO there are many more signals to predict especially when the number of microphones is large.
However, after including the second network, MIMO+ MVDR+WPE+MISO$_5$ produces a performance competitive to MISO$_1$+MVDR+WPE+MISO$_4$.

We shall point out that in \cite{Han2020}, published in the same conference as our preliminary study \cite{Wang2020d}, a MIMO network based on Conv-TasNet is proposed for binaural speaker separation.
Their contribution is on preserving spatial awareness, while ours is on the reduction of computation, and MIMO's integration with WPE, beamforming, and post-filtering. %

\subsection{MISO$_1$+mMVDR+WPE+MISO$_6$ System}

\begin{figure}
  \centering  
  \includegraphics[width=8.5cm]{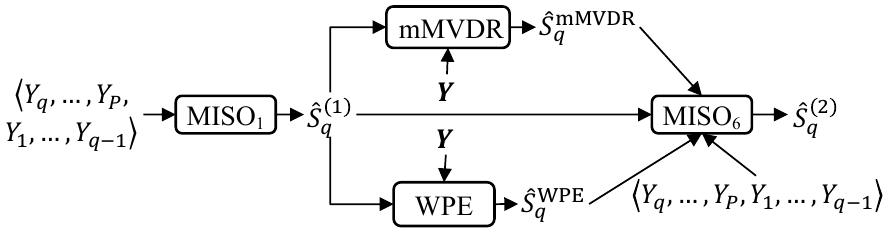}
  \caption{MISO$_1$+mMVDR+WPE+MISO$_6$ system.}
  \label{MISO1+mMVDR+WPE+MISO6}\vspace{-0.4cm}
\end{figure}

In the previous subsections, we predict the target speech at all the microphones using MISO$_1$ or MIMO in order to compute an MVDR beamformer or a WPE filter.
However, to compute them, we do not have to estimate the target speech at all the microphones, and can instead derive a mask-based MVDR (mMVDR) beamformer \cite{Yoshioka2015, Taherian2020} that only uses the target speech estimate at the reference microphone.
Fig.~\ref{MISO1+mMVDR+WPE+MISO6} illustrates our proposed system.
Based on the output of MISO$_1$ at the reference microphone, we compute a real-valued T-F mask, and use it to compute target and non-target covariance matrices \cite{Yoshioka2015, Heymann2015, Taherian2020}:
\begin{align}
\hat{\mathbf{\Phi}}^{(s)}(f) &= \sum\nolimits_t \hat{m}(t,f) \hat{\mathbf{Y}}(t,f)\hat{\mathbf{Y}}(t,f)^{\H} \label{covariancematrixMasking} \\
\hat{\mathbf{\Phi}}^{(v)}(f) &= \sum\nolimits_t \big(1-\hat{m}(t,f)\big) \hat{\mathbf{Y}}(t,f)\hat{\mathbf{Y}}(t,f)^{\H} \\
\hat{m}(t,f) &= \frac{|\hat{S}_q^{(1)}(t,f)|}{|\hat{S}_q^{(1)}(t,f)| + |\hat{Y}_q(t,f) - \hat{S}_q^{(1)}(t,f)|}.
\end{align}
An MVDR beamformer is then computed using Eqs.~(\ref{steeringvec}) and (\ref{mvdrfilter}), and the beamforming result is obtained using (\ref{mvdrresult}).

The WPE filter is computed using Eq.~(\ref{dnnwpe}), but we compute the denominator as
\begin{align}\label{lambdasingleestimate}
\hat{\lambda}(t,f) =\text{max}\big(\varepsilon \text{max}(|\hat{S}_q^{(1)}|^2), |\hat{S}_q^{(1)}(t,f)|^2\big),
\end{align}
where, different from (\ref{lambdaallestimates}), $\hat{\lambda}$ is not computed by summing the target estimates over all the microphones.
The dereverberation result $\hat{S}_q^{\text{WPE}}$ is computed using Eq.~(\ref{wperesult}).

This system can use MISO$_1$'s output, which is expected to be better than MIMO's output, for beamforming and WPE. In addition, the first network runs only once at run time and the system also avoids the reliance on uniform circular geometry. 

\subsection{MISO$_1$+mMVDR\_WPE+WPE+MISO$_7$ System}\label{cascadeWPEandBF}

The above systems stack the beamforming and the WPE results as extra inputs to the second network.
Following \cite{Drude2018}, the beamforming filter can be computed to filter the WPE results (i.e., $\hat{\mathbf{S}}^{\text{WPE}}(t,f)=\big[\hat{S}_1^{\text{WPE}}(t,f), \dots, \hat{S}_P^{\text{WPE}}(t,f)\big]^{\T}$) rather than the mixture.
The rationale is that if the mixture becomes less reverberant after being processed by WPE, estimated steering vectors and beamforming results would be better.
Cascading WPE and beamforming is a popular technique in the REVERB and CHiME challenges \cite{Kinoshita2017, Boeddecker2018} for robust ASR.

\begin{figure}
  \centering  
  \includegraphics[width=8.5cm]{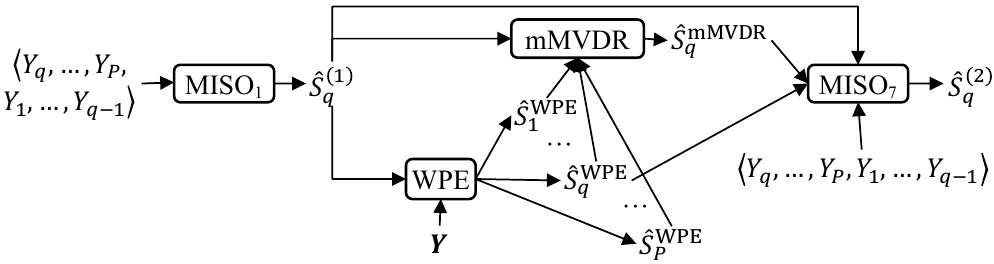}
  \caption{MISO$_1$+mMVDR\_WPE+WPE+MISO$_7$ system.}
  \label{MISO1+mMVDR_WPE+WPE+MISO7}\vspace{-0.4cm}
\end{figure}

Fig.~\ref{MISO1+mMVDR_WPE+WPE+MISO7} illustrates this system.
The target and non-target covariance matrices are computed as
\begin{align}
\hat{\mathbf{\Phi}}^{(s)}(f) &= \sum\nolimits_t \hat{m}(t,f) \hat{\mathbf{S}}^{\text{WPE}}(t,f)\hat{\mathbf{S}}^{\text{WPE}}(t,f)^{\H}, \label{covariancematrixMaskSph} \\
\hat{\mathbf{\Phi}}^{(v)}(f) &= \sum\nolimits_t \big(1-\hat{m}(t,f)\big) \hat{\mathbf{S}}^{\text{WPE}}(t,f)\hat{\mathbf{S}}^{\text{WPE}}(t,f)^{\H}, \label{covariancematrixMaskNoi} \\
\hat{m}(t,f) &= \frac{|\hat{S}_q^{(1)}(t,f)|}{|\hat{S}_q^{(1)}(t,f)| + |\hat{S}_q^{\text{WPE}}(t,f) - \hat{S}_q^{(1)}(t,f)|}. \label{wpemask}
\end{align}
An MVDR beamformer $\hat{\mathbf{w}}(f;q)$ is then computed using Eqs.~(\ref{steeringvec}) and (\ref{mvdrfilter}), and the beamforming result is obtained as
\begin{align}
\hat{S}_q^{\text{mMVDR}}(t,f) = \hat{\mathbf{w}}(f;q)^{\H}\hat{\mathbf{S}}^{\text{WPE}}(t,f).
\end{align}

This system is denoted as MISO$_1$+mMVDR\_WPE+WPE +MISO$_7$, where mMVDR\_WPE means that mMVDR is applied to WPE results.
This naming convention also applies to our subsequent systems.
We emphasize that mMVDR\_WPE is expected to produce low-distortion results.
It can likely better suppress non-target signals than WPE or mMVDR alone, and therefore its results could serve as a better feature to the second network, according to our analysis in the introduction.

\subsection{MISO$_1$+mWMPDR\_WPE+WPE+MISO$_8$ System}\label{convolutionalbeamformer}

The recently-proposed convolutional beamformer \cite{Nakatani2019ConvBeamformer, Nakatani2020, ZhangWY2021} has shown strong performance in speech dereverberation and separation, and robust ASR.
A subsequent study \cite{Boeddeker2020} proves that a convolutional beamformer can be factorized into a product of a WPE filter and a weighted minimum power distortionless response (WMPDR) beamformer.
Our study leverages them to obtain low-distortion target estimates.

\begin{figure}
  \centering  
  \includegraphics[width=8.5cm]{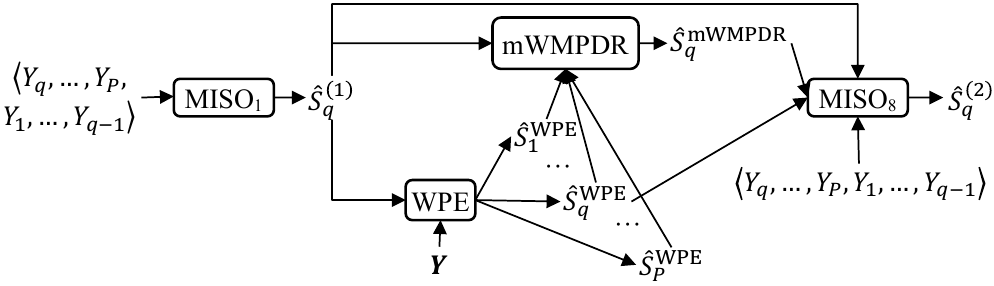}
  \caption{MISO$_1$+mWMPDR\_WPE+WPE+MISO$_8$ system.}
  \label{MISO1+mWMPDR_WPE+WPE+MISO8}\vspace{-0.4cm}
\end{figure}

Fig.~\ref{MISO1+mWMPDR_WPE+WPE+MISO8} depicts the system.
The WPE filter is obtained with $\hat{\lambda}$ computed using Eq.~(\ref{lambdasingleestimate}).
The mask-based WMPDR (mWMPDR) beamformer is computed as
\begin{align}\label{wmpdrfilter}
\hat{\mathbf{w}}(f;q) = \frac{\hat{\mathbf{\Phi}}^{(y')}(f)^{-1} \hat{\mathbf{d}}(f)}{\hat{\mathbf{d}}(f)^{\H} \hat{\mathbf{\Phi}}^{(y')}(f)^{-1} \hat{\mathbf{d}}(f)} \hat{d}_q^{*}(f),
\end{align}
where $\hat{\mathbf{d}}(f)$ is the principal eigenvector extracted from $\hat{\mathbf{\Phi}}^{(s)}(f)$ computed in Eq.~(\ref{covariancematrixMaskSph}), and $\hat{\mathbf{\Phi}}^{(y')}(f)$ is computed as
\begin{align}\label{covariancematrixMWPDR}
\hat{\mathbf{\Phi}}^{(y')}(f) &= \sum\nolimits_t \frac{\hat{\mathbf{S}}^{\text{WPE}}(t,f)\hat{\mathbf{S}}^{\text{WPE}}(t,f)^{\H}}{\hat{\lambda}(t,f)}
\end{align}
with $\hat{\lambda}$ computed using (\ref{lambdasingleestimate}).
The beamforming result is
\begin{align}\label{mWMPDRresult}
\hat{S}_q^{\text{mWMPDR}}(t,f) = \hat{\mathbf{w}}(f;q)^{\H}\hat{\mathbf{S}}^{\text{WPE}}(t,f).
\end{align}

Both filters assume that the target speech follows a complex Gaussian distribution with a time-varying PSD.
The two filters are jointly optimal in the sense that they can maximize the likelihood of the target speech under the hypothesized Gaussian distribution \cite{Nakatani2019ConvBeamformer, Nakatani2020}.
The PSD $\hat{\lambda}$ is computed based on DNN outputs, following \cite{Kinoshita2017, Nakatani2020}.
Notice that here we use the factorized solution \cite{Boeddeker2020}.
This way, we can include the intermediate WPE result to train the second network, and compute the steering vector based on the WPE result using Eqs.~(\ref{steeringvec}) and (\ref{covariancematrixMaskSph}).
This produces better performance than the non-factorized solution, which does not provide the intermediate WPE result, and needs to compute the steering vector based on Eq.~(\ref{covariancematrixMasking}), which is usually worse as the mixture outer product in (\ref{covariancematrixMasking}) is typically less accurate than that in (\ref{covariancematrixMaskSph}).

\begin{figure}
  \centering  
  \includegraphics[width=8.5cm]{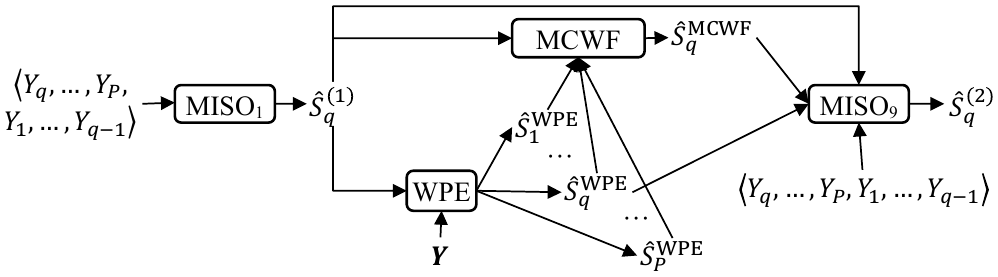}
  \caption{MISO$_1$+MCWF\_WPE+WPE+MISO$_9$ system.}
  \label{MISO1+MCWF_WPE+WPE+MISO9}\vspace{-0.4cm}
\end{figure}

\subsection{MISO$_1$+MCWF\_WPE+WPE+MISO$_9$ System}

Since MISO$_1$ is designed to only provide $\hat{S}_q^{(1)}$, the above systems use T-F masks to compute covariance matrices for MVDR or MPDR beamforming.
A simpler way to perform beamforming is to compute a multi-channel Wiener filter (MCWF) $\mathbf{w}(f;q) \in \CC^{P}$ by filtering the WPE results $\hat{\mathbf{S}}^{\text{WPE}}$ to approximate $\hat{S}_q^{(1)}$.
The minimization problem is
\begin{align}\label{mcwfonwpe}
\underset{\mathbf{w}(f;q)}{{\text{argmin}}} \sum\nolimits_{t} |\hat{S}_q^{(1)}(t,f)-\mathbf{w}(f;q)^{\H}\ \hat{\mathbf{S}}^{\text{WPE}}(t,f)|^2,
\end{align}
and the beamforming result is 
\begin{align}\label{mcwfonwperesult}
\hat{S}_q^{\text{MCWF}}(t,f) = \hat{\mathbf{w}}(f;q)^{\H}\hat{\mathbf{S}}^{\text{WPE}}(t,f),
\end{align}
where $\hat{\mathbf{S}}^{\text{WPE}}$ is obtained with $\hat{\lambda}$ computed using Eq.~(\ref{lambdasingleestimate}).
This way, DNN-estimated phase can be utilized for beamforming.
This system is illustrated in Fig.~\ref{MISO1+MCWF_WPE+WPE+MISO9}.

\subsection{Monaural Systems}

\begin{figure}
  \centering  
  \includegraphics[width=6cm]{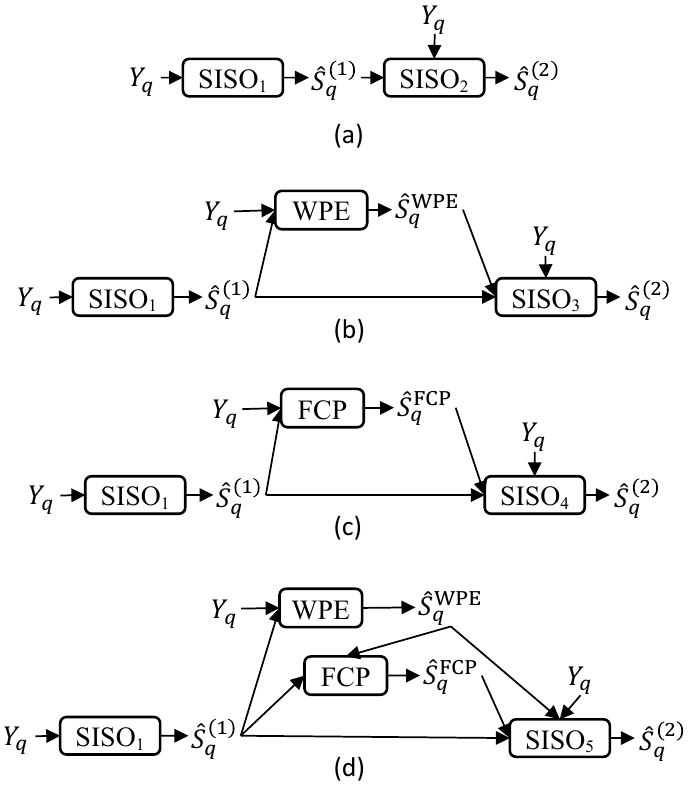}
  \caption{
  (a) SISO$_1$+SISO$_2$;
  (b) SISO$_1$+WPE+SISO$_3$;
  (c) SISO$_1$+FCP+ SISO$_4$;
  (d) SISO$_1$+FCP\_WPE+WPE+SISO$_5$ systems.}
  \label{single_channel_models}\vspace{-0.3cm}
\end{figure}

All the above systems are multi-channel.
For monaural processing, our baselines are the SISO$_1$ and SISO$_1$+SISO$_2$ systems shown in Fig.~\ref{single_channel_models}(a), where SISO means \textit{single-microphone input and single-microphone output}.
We can use monaural WPE to get low-distortion target estimates.
The resulting system, SISO$_1$+WPE+SISO$_3$, is shown in Fig.~\ref{single_channel_models}(b).
We can also use our recently-proposed monaural FCP algorithm \cite{Wang2021FCPjournal} for dereverberation.
The resulting system, SISO$_1$+FCP+SISO$_4$, is shown in Fig.~\ref{single_channel_models}(c).
Compared with monaural WPE, FCP was shown to better improve the second network in a recent 2stage-DNN system and better reduce early reflections \cite{Wang2021FCPjournal}.
The FCP filter $\hat{\mathbf{g}}'(f)$ is obtained by solving the following problem:
\begin{align}\label{1chFCP}
\underset{\mathbf{g}'(f)}{{\text{argmin}}} \sum_{t} \frac{|Y_q(t,f)-\mathbf{g}'(f)^{\H}\ \widetilde{\hat{\mathbf{S}}}{}_q^{(1)}(t,f)|^2}{\hat{\eta}(t,f)},
\end{align}
where $\mathbf{g}'(f)\in \CC^{K'}$ is a $K'$-dimensional filter and $\widetilde{\hat{\mathbf{S}}}{}_q^{(1)}=[\hat{S}_q^{(1)}(t,f),\dots,\hat{S}_q^{(1)}(t-K'+1,f)]^\T$. 
Slightly different from that in \cite{Wang2021FCPjournal}, $\hat{\eta}(t,f)$ is here defined as
\begin{align}\label{FCPweight}
\hat{\eta}(t,f) =\text{max}(\varepsilon' \text{max}(|Y_q - \hat{S}_q^{(1)}|^2), |Y_q(t,f) - \hat{S}_q^{(1)}(t,f)|^2).
\end{align}
Given that $Y_q=X_q+N_q$ and assuming that the reverberant speech $X_q$ and the target direct-path signal estimate $\hat{S}_q^{(1)}$ are uncorrelated with the noise $N_q$, FCP forwardly filters $\hat{S}_q^{(1)}$ to approximate $X_q$ contained in $Y_q$ \cite{Wang2021FCPjournal}.
As a result, $\hat{\mathbf{g}}'(f)^{\H}\ \widetilde{\hat{\mathbf{S}}}{}_q^{(1)}(t,f)$ should be an estimate of $X_q(t,f)$, and $\hat{\mathbf{g}}'(f)^{\H}\ \widetilde{\hat{\mathbf{S}}}{}_q^{(1)}(t,f) - \hat{S}_q^{(1)}(t,f)$ an estimate of the reverberation of the target speaker.
The FCP result is obtained by subtracting the estimated reverberation from the mixture:
\begin{align}\label{FCPresult}
\hat{S}_q^{\text{FCP}}(t,f) = Y_q(t,f) - \Big( \hat{\mathbf{g}}'(f)^{\H}\ \widetilde{\hat{\mathbf{S}}}{}_q^{(1)}(t,f) - \hat{S}_q^{(1)}(t,f) \Big).
\end{align}

Note that SISO$_1$+WPE+SISO$_3$ and SISO$_1$+FCP+SISO$_4$ have been proposed in our recent study \cite{Wang2021FCPjournal}.
Our new contribution is performing FCP on the WPE result.
Fig.~\ref{single_channel_models}(d) illustrates the idea.
The FCP result is computed by replacing $Y_q$ in Eqs.~(\ref{1chFCP})-(\ref{FCPresult}) with the monaural WPE result $\hat{S}_q^{\text{WPE}}$.
This way, we can combine WPE's ability at producing low-distortion estimates, and FCP's abilities at reducing early reflections and aggressively suppressing the reverberation of the target speaker.

\subsection{MISO$_1$+FCP\_mWMPDR\_WPE+mWMPDR\_WPE+WPE+ MISO$_{10}$ System}

\begin{figure}
  \centering  
  \includegraphics[width=8.5cm]{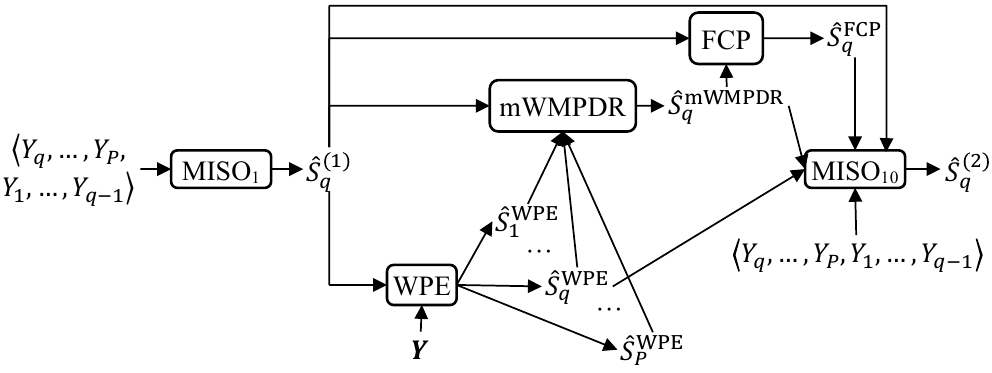}
  \caption{MISO$_1$+FCP\_mWMPDR\_WPE+mWMPDR\_WPE+WPE+MISO$_{10}$ system.}
  \label{FCP+cb}\vspace{-0.3cm}
\end{figure}

Building upon MISO$_1$+mWMPDR\_WPE+WPE+MISO$_8$ in Section~\ref{convolutionalbeamformer}, this system (see Fig.~\ref{FCP+cb})
applies a monaural FCP filter to the output of mWMPDR\_WPE for further dereverberation.
The FCP filter is computed by replacing $Y_q$ in Eqs.~(\ref{1chFCP})-(\ref{FCPresult}) with $\hat{S}_q^{\text{mWMPDR}}$.
Besides the motivation behind combining monaural FCP and WPE in the previous subsection, another motivation is that the WPE module has a filter length of $KP$, and, as the number of microphones $P$ gets large, the number of filter taps $K$ is usually decreased to avoid introducing extra amount of computation.
For example, in \cite{Boeddeker2020}, $K$ is set to 37 when $P$ is 1, and to 10 when $P$ is 8.
However, a smaller $K$ for a larger $P$ limits the amount of contextual information.
The additional monaural filter can use more filter taps without introducing a large amount of computation.

\section{DNN Configurations}\label{dnndescription}

\subsection{Complex Spectral Mapping}\label{complexmappingdescription}

Our DNNs are trained to predict the RI components of the target direct-path signal from the mixture RI components.
This approach and the related complex ratio masking technique \cite{Williamson2016} have shown strong performance in speech separation \cite{Fu2017, Liu2019, Tan2020CSM, Wang2020b, Wang2020d, Wang2020a, Wang2020c}.
In our experiments, complex spectral mapping shows clearly better performance than complex ratio masking.
Following \cite{Wang2020d, Wang2021compensation}, for MISO and SISO networks the loss is defined on the predicted RI components and their magnitude:
\begin{multline}
\mathcal{L}^{(b)}_{q,\text{RI+Mag}} = \Big\| \hat{R}_q^{(b)} - \text{Real}(S_q)\Big\|_1 + \Big\| \hat{I}_q^{(b)} - \text{Imag}(S_q)\Big\|_1
 \\
+ \Big\| \sqrt{{\hat{R}_q^{{(b)}^2}}+{\hat{I}_q^{{(b)}^2}}} - |S_q|\Big\|_1 ,\label{enhlossMISO}
\end{multline}
where $\hat{R}^{(b)}$ and $\hat{I}^{(b)}$ are the estimated RI components produced by using linear activation in the output layer, $b\in\{1,2\}$ denotes which one of the two DNNs produces the estimates, $\text{Real}(\cdot)$ and $\text{Imag}(\cdot)$ extract RI components, and $\| \cdot\|_1$ computes the $L_1$ norm.
The enhancement result is obtained as $\hat{S}_q^{(b)}=\hat{R}_q^{(b)}+j\hat{I}_q^{(b)}$.
Inverse STFT is applied for signal re-synthesis.
For the MIMO network, the loss is $\mathcal{L}^{(1)}_{\text{RI+Mag}} = \sum\nolimits_{p=1}^P \mathcal{L}^{(1)}_{p,\text{RI+Mag}}$.
We only use superscript $(1)$ here, as MIMO is only used as the first network.

\begin{figure}
  \centering  
  \includegraphics[width=8cm]{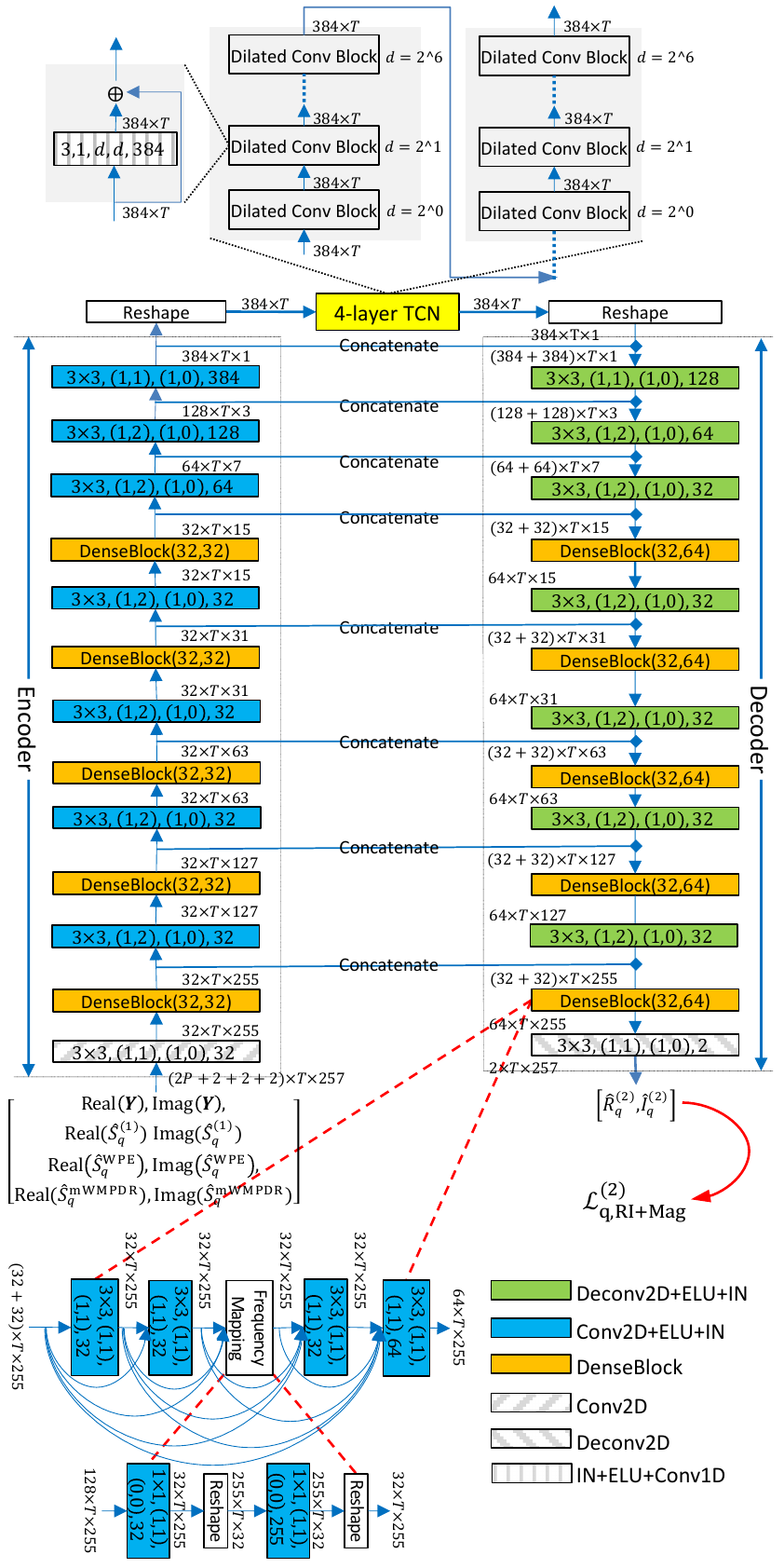}
  \caption{Network architecture of MISO$_8$.
  The tensor shape after each encoder-decoder block is in the format: \textit{featureMaps$\times$timeSteps$\times$frequencyChannels}. Each one of Conv2D$+$ELU$+$IN, Deconv2D$+$ELU$+$IN, Conv2D, and Deconv2D blocks is shown in the format: \textit{kernelSizeTime$\times$kernelSizeFreq}, \textit{(stridesTime, stridesFreq)}, \textit{(paddingsTime, paddingsFreq)}, \textit{featureMaps}.
  Each DenseBlock($g_1$,$g_2$) has five Conv2D+ELU+IN blocks with growth rate $g_1$ for the first four layers and $g_2$ for the last one.
  The tensor shape after each TCN block is in the format \textit{featureMaps$\times$timeSteps}.
  Each IN$+$ELU$+$Conv1D block is specified in the format: \textit{kernelSizeTime, stridesTime, paddingsTime, dilationTime, featureMaps}.}
  \label{dnnfigure}
  \vspace{-0.6cm}
\end{figure}

\subsection{DNN Architecture}\label{dnnarchitectures}

As an example, the network architecture of MISO$_8$ in MISO$_1$+mWMPDR\_WPE+WPE+MISO$_8$ is shown in Fig.~\ref{dnnfigure}.
Other networks use the same architecture but differ in the network input and output, as they use different signals there. %
This architecture follows recent studies in complex T-F domain speech separation \cite{Liu2019, Wang2020a, Wang2020b, Wang2020c, Wang2020d, Tan2021}.
It is a temporal convolutional network (TCN) sandwiched by a U-Net.
DenseNet blocks are inserted at multiple frequency scales in the encoder and decoder of the U-Net.
The encoder contains one two-dimensional (2D) convolution, and seven convolutional blocks, each with 2D convolution, exponential linear units (ELU) non-linearity, and instance normalization (IN), for down-sampling.
The decoder includes seven blocks of 2D deconvolution, ELU, and IN, and one 2D deconvolution, for up-sampling.
The TCN contains four layers, each with seven dilated convolutional blocks.
One one-dimensional (1D) depth-wise separable convolution is used in each dilated convolutional block. %
The RI components of different input/output signals are stacked along the feature map dimension for the network input/output.

\section{Experimental Setup}\label{setup}

We evaluate the proposed algorithms on two tasks: speech dereverberation with air-conditioning noise, and speech enhancement in noisy-reverberant conditions with challenging non-stationary noise.
This section describes the dataset used for each task, the hyper-parameter settings, the evaluation metrics, and the baseline systems.

\subsection{Dataset for Speech Dereverberation}

For speech dereverberation, we train our models on a simulated reverberant dataset with air-conditioning noise.
Besides evaluating the trained models on the simulated data, we apply them, without re-training, to the REVERB corpus \cite{Kinoshita2016} to show their effectiveness at dealing with real reverberant recordings.

We use the clean signals in the WSJCAM0 corpus for simulation.
It contains 7,861, 742, and 1,088 utterances respectively in its training, validation, and test sets.
We use them to respectively simulate 39,305 (7,861$\times$5), 2,968 (742$\times$4), and 3,264 (1,088$\times$3) noisy-reverberant mixtures as our training, validation, and test sets.
The spatialization procedure follows our previous study \cite{Wang2020d}, where, for each utterance, we randomly sample a room with random room characteristics and speaker and microphone locations, using the \textit{pyroomacoustics} RIR generator.
For each utterance, we add a diffuse air-conditioning noise drawn from the REVERB corpus \cite{Kinoshita2016}.
The SNR between the target anechoic speech and the noise is sampled from the range $[5, 25]$ dB.
The speaker-to-microphone distance is sampled from the range $[0.75, 2.5]$ m.
The simulated array is an eight-microphone uniform circular array with a 10 cm radius.
The reverberation time (T60) is drawn from the range $[0.2, 1.3]$ s.
The sampling rate is 16 kHz.

We apply the trained models directly to the ASR tasks of REVERB.
The test mixtures are recorded using an array with the same geometry in real rooms with T60 approximately 0.7 s and with speaker-to-microphone distance around 1 m in the near-field case and 2.5 m in the far-field case.
The recorded mixtures contain weak air-conditioning noise.

The most recent Kaldi recipe is used to build our ASR backend.
It is trained using the official noisy-reverberant speech plus dry source signals of REVERB.
We feed enhanced time-domain signals to the backend for decoding.

\subsection{Dataset for Noisy-Reverberant Speech Enhancement}

The clean signals for simulation are also from WSJCAM0.
We use them to simulate 39,245, 2,965, and 3,260 noisy-reverberant mixtures as our training, validation, and test sets, respectively.
We generate noises by using the FSD50k dataset \cite{Fonseca2020}, which contains around 50,000 Freesound clips with human-labeled sound events distributed in 200 classes drawn from the AudioSet ontology.
We use the clips in the development set of FSD50k to simulate the noises for training and validation, and those in the evaluation set to simulate the noises for testing.
Since our task is speech enhancement, following \cite{Tzinis2021}, we remove the clips containing any sound produced by humans, based on the sound event annotations.
Such clips have annotations such as \textit{Human\_voice}, \textit{Male\_speech\_and\_man\_speaking}, \textit{Yell},
\textit{Giggle}, etc.
To generate multi-channel noise signals, we sample up to seven noise clips for each mixture.
We treat each sampled clip as a point source in the space, convolve each source with the corresponding RIR, and add the convolved signals together to create the mixture.
The directions of each noise source and the target speaker to the array are independently sampled from the range $[0,2\pi]$.
Following the setup in the FUSS dataset \cite{Wisdom2021}, which is designed for universal sound separation, we consider noise clips as background noises if they are longer than ten seconds and as foreground noises otherwise, and each simulated mixed noise file has one background noise and the rest are foreground noises.
The energy level between the dry background noise and each dry foreground noise is sampled from the range $[-3, 9]$ dB.
Considering that some FSD50k clips contain digital zeros, silence or transient sounds, the energy level is computed by first removing silent segments in each clip, then computing a sample variance based on the remaining samples, and then scaling the clips to an SNR based on the sample variance of each clip.
After summing up all the reverberant noises, we scale the summated reverberant noises such that the SNR between the target anechoic speech and the summated reverberant noises is equal to a value sampled from the range $[-8, 3]$ dB.
In addition to the FSD50k clips, in each mixture we always add a very weak, diffuse, stationary air-conditioning noise drawn from REVERB, where the SNR between the target anechoic speech and the noise is sampled from the range $[10, 30]$ dB.
The distance between each source and the array center is sampled from the range $[0.75, 2.5]$ m.
The T60 is drawn from the range $[0.2, 1.0]$ s.
The simulated array is a six-microphone uniform circular array with a 10 cm radius.
The sampling rate is 16 kHz.

\begin{table*}[t]
\scriptsize
\centering
\captionsetup{justification=centering}
\sisetup{table-format=2.2,round-mode=places,round-precision=2,table-number-alignment = center,detect-weight=true,detect-inline-weight=math}
\caption{\textsc{One- and Eight-Microphone SI-SDR (dB), PESQ, eSTOI (\%), PDSAcc (\%), and pSNR (dB) Results \\ on The Speech Dereverberation Task, and WER (\%) on Real Data of REVERB.}}
\label{resultsspeechdereverb}
\setlength{\tabcolsep}{4pt}
\begin{tabular}{clcS[table-format=2.1,round-precision=1]S[table-format=1.1]!{\enspace}S[table-format=2.1,round-precision=1]S[table-format=2.1,round-precision=1]S[table-format=2.1,round-precision=1]!{\enspace}SSS!{\enspace}SSS}
\toprule
& & \multirow{2}{*}[-6.5pt]{\begin{tabular}[c]{@{}c@{}}Run $\text{DNN}_1$ \\ \#times\end{tabular}} & & & & & & \multicolumn{3}{c}{WER on val. set} & \multicolumn{3}{c}{WER on test set} \\
\cmidrule(lr{9pt}){9-11} \cmidrule(lr){12-14}
\multicolumn{1}{l}{Entry} & \multicolumn{1}{l}{Systems} & & \multicolumn{1}{c}{SI-SDR} & {PESQ}\hspace{-.5em} & \hspace{-.2em}eSTOI & {PDSAcc} & {pSNR} & \multicolumn{1}{c}{Near} & \multicolumn{1}{c}{Far} & {Avg.} & \multicolumn{1}{c}{Near} & \multicolumn{1}{c}{Far} & {Avg.} \\

\midrule

0a & Unprocessed & - & -3.6 & 1.64 & 49.4 & {-} & -3.7 & 15.35 & 16.88 & 16.11 & 17.09 & 17.29 & 17.19 \\

\midrule

1a & SISO$_1$ & 1 & 8.4 & 3.12 & 86.8 & 70.7 & 10.2 & 8.61 & 10.18 & 9.39 & 8.97 & 9.49 & 9.23 \\
1b & SISO$_1$+SISO$_2$ & 1 & 9.0 & 3.17 & 87.7 & 71.9 & 10.9 & 8.98 & 10.46 & 9.72 & 8.43 & 9.15 & 8.79 \\
1c & SISO$_1$+WPE+SISO$_3$ & 1 & 11.4 & 3.39 & 90.5 & 76.9 & 13.5 & 8.80 & 10.73 & 9.77 & 8.18 & 9.35 & 8.77 \\
1d & SISO$_1$+FCP+SISO$_4$ & 1 & 12.0 & 3.47 & 91.4 & 77.6 & 14.1 & 8.67 & 10.32 & 9.50 & 7.70 & 8.20 & 7.95 \\
1e & SISO$_1$+FCP\_WPE+WPE+SISO$_5$ & 1 & 12.7 & 3.49 & 91.9 & 79.0 & 14.9 & 8.23 & 10.39 & 9.31 & 7.54 & 7.97 & 7.75 \\
\hdashline
1f & SISO$_1$+WPE & 1 & -1.6 & 1.83 & 61.2 & 58.0 & 3.3 & 13.79 & 14.35 & 14.07 & 12.78 & 14.62 & 13.70 \\
1g & SISO$_1$+FCP & 1 & 2.8 & 1.84 & 62.3 &  67.4 & 8.8 & 16.84 & 19.96 & 18.40 & 16.22 & 15.90 & 16.06 \\
1h & SISO$_1$+FCP\_WPE & 1 & 4.4 & 1.88 & 66.4 &  67.3 & 8.8 & 17.53 & 19.89 & 18.71 & 14.63 & 15.90 & 15.27 \\

\midrule

2a & MISO$_1$ & 1 & 11.3 & 3.49 & 92.1 & 76.2 & 12.9 & 9.23 & 9.16 & 9.20 & 6.71 & 7.19 & 6.95 \\
2b & MISO$_1$+MISO$_3$ & 1 & 11.8 & 3.60 & 92.6 & 77.3 & 13.5 & 9.11 & 9.09 & 9.10 & 6.39 & 6.82 & 6.61 \\
2c & MISO$_1$+MVDR+MISO$_2$ & $P$ & 14.4 & 3.73 & 94.2 & 83.3 & 16.7 & 8.73 & \bfseries 7.86 & 8.29 & 5.78 & 7.02 & 6.40 \\
2d & MISO$_1$+MVDR+WPE+MISO$_4$ & $P$ & 16.3 & 3.82 & 95.2 & 85.8 & 18.6 & 8.23 & 8.20 & 8.21 & 5.91 & 6.92 & 6.42 \\
\hdashline
2e & MISO$_1$+MVDR & $P$ & 7.7 & 2.19 & 77.7 & 77.7 & 11.3 & 9.67 & 11.28 & 10.47 & 7.63 & 8.64 & 8.13 \\
2f & MISO$_1$+WPE & $P$ & 3.1 & 2.26 & 76.7 & 67.3 & 5.8 & 12.35 & 14.56 & 13.46 & 10.60 & 11.24 & 10.92 \\

\midrule

3a & MIMO & 1 & 9.5 & 3.33 & 90.8 & 72.4 & 11.0 & 8.98 & 11.14 & 10.06 & 6.96 & 7.26 & 7.11 \\
3b & MIMO+MVDR+WPE+MISO$_5$ & 1 & 16.1 & 3.80 & 95.0 & 85.3 & 18.3 & \bfseries 7.74 & 8.82 & 8.28 & 5.59 & \bfseries 6.31 & 5.95 \\
\hdashline
3c & MIMO+MVDR & 1 & 6.0 & 2.18 & 77.6 & 76.2 & 10.3 & 10.61 & 11.69 & 11.15 & 7.57 & 8.41 & 7.99 \\
3d & MIMO+WPE & 1 & 3.2 & 2.25 & 76.8 & 67.3 & 5.9 & 14.72 & 16.88 & 15.80 & 11.11 & 13.07 & 12.09 \\

\midrule

4a & MISO$_1$+mMVDR+WPE+MISO$_6$ & 1 & 16.0 & 3.80 & 95.1 & 85.5 & 18.2 & 8.48 & 8.07 & 8.28 & 5.88 & 7.02 & 6.45 \\
4b & MISO$_1$+mMVDR\_WPE+WPE+MISO$_7$ & 1 & 17.2 & 3.91 & 96.0 & 87.4 & 19.5 &
7.92 & 8.20 & \bfseries 8.06 & 6.23 & 6.72 & 6.47 \\
4c & MISO$_1$+mWMPDR\_WPE+WPE+MISO$_8$ & 1 & 17.8 & 3.96 & 96.5 & 88.2 & 20.1 &
8.23 & 8.34 & 8.29 & 5.84 & 6.89 & 6.37 \\
4d & MISO$_1$+MCWF\_WPE+WPE+MISO$_9$ & 1 & 17.2 & 3.94 & 96.3 & 87.4 & 19.6 & 8.42 & 8.00 & 8.21 & 5.81 & 6.62 & 6.21 \\
\hdashline
4f & MISO$_1$+WPE & 1 & 3.1 & 2.26 & 76.8 & 67.3 & 5.9 & 14.72 & 17.09 & 15.91 & 11.40 & 12.83 & 12.12 \\
4g & MISO$_1$+mMVDR & 1 & 1.8 & 2.23 & 75.1 & 69.8 & 5.5 & 11.67 & 10.80 & 11.23 & 7.89 & 8.81 & 8.35 \\
4h & MISO$_1$+mMVDR\_WPE & 1 & 5.4 & 2.76 & 86.5 & 73.9 & 7.2 & 11.92 & 12.92 & 12.42 & 8.40 & 9.62 & 9.01 \\
4i & MISO$_1$+mWMPDR\_WPE & 1 & 5.6 & 2.95 & 88.7 & 75.7 & 7.5 & 11.60 & 12.85 & 12.22 & 7.44 & 9.39 & 8.42 \\
4j & MISO$_1$+MCWF\_WPE & 1 & 14.6 & 2.85 & 91.9 & 82.9 & 16.3 & 10.36 & 11.41 & 10.88 & 7.09 & 8.71 & 7.90 \\

\midrule

5a & \begin{tabular}{@{}l@{}}MISO$_1$+FCP\_mWMPDR\_WPE+ \\ \,\,\,\,\,\,\,\,\,\,\,\,mWMPDR\_WPE+WPE+MISO$_{10}$\end{tabular}
 & 1 & \bfseries 18.2 & \bfseries 3.98 & \bfseries 96.7 & \bfseries 88.6 & \bfseries 20.5 & 8.23 & 8.27 & 8.25 & 5.59 & 6.85 & 6.22 \\
\hdashline
5b & MISO$_1$+FCP\_mWMPDR\_WPE & 1 & 12.6 & 2.88 & 91.3 & 81.3 & 14.6 & 10.54 & 10.59 & 10.56 & 6.96 & 8.34 & 7.65 \\

\midrule

6a & MISO$_1$+GEV+MISO$_{2a}$ & $P$ & 14.1 & 3.71 & 93.9 & 82.6 & 16.2 & 8.73 & 8.75 & 8.74 & 5.97 & 6.79 & 6.38 \\ %
6b & MISO$_1$+MCWF+MISO$_{2b}$ & $P$ & 14.9 & 3.80 & 94.9 & 84.2 & 17.0 & 7.99 & 8.68 & 8.34 & \bfseries 5.49 & 6.55 & \bfseries 6.02 \\
\hdashline
6c & MISO$_1$+GEV & $P$ & -9.2 & 2.14 & 72.1 & 68.3 & -1.3 & 11.29 & 12.03 & 11.66 & 8.02 & 9.79 & 8.90 \\
6d & MISO$_1$+MCWF & $P$ & 10.0 & 2.31 & 82.1 & 79.1 & 13.2 & 9.61 & 10.39 & 10.00 & 7.41 & 8.37 & 7.89 \\

\midrule

7a & Oracle spectral magnitude mask & - & 1.7 & 3.42 & 91.6 & {-} & {-} & {-} & {-} & {-} & {-} & {-} & {-} \\
7b & Oracle phase-sensitive mask & - & 6.3 & 3.62 & 91.3 & {-} & {-} & {-} & {-} & {-} & {-} & {-} & {-} \\

\bottomrule
\end{tabular}\vspace{-0.4cm}
\end{table*}

\subsection{Miscellaneous Configurations}

We consider 1- and 8-channel processing for the dereverberation task, and 1-, 2-, and 6-channel processing for the enhancement task.
For the 2-channel setup, we use the first and the fourth microphone signals as input.
The first microphone is always considered as the reference microphone.
Although both tasks use uniform circular arrays, it should be obvious that the proposed algorithms are not limited to this array geometry.

For STFT, the window size is 32 ms, the hop size is 8 ms, and the analysis window is the square root of the Hann window.
A 512-point fast Fourier transform is applied to extract 257-dimensional STFT spectrums. 
We normalize the sample variance of each mixture to one before any processing.
During training, the target signal is scaled by the same factor used for scaling the mixture.
No sentence- or global-level mean-variance normalization is performed on input features. 

For DNN-WPE, through cross-validation we respectively set the number of filter taps $K$ to 37, 30, 10 and 8 in 1-, 2-, 6- and 8-microphone cases, the prediction delay $\Delta$ to 3, following \cite{Kinoshita2016, Kinoshita2017}, and
$\varepsilon$ in Eqs.~(\ref{lambdaallestimates}) and (\ref{lambdasingleestimate}) to $1e^{-5}$.
For FCP, $K'$ is set to 40, and $\varepsilon'$ in Eq.~(\ref{FCPweight}) to $1e^{-3}$.

The two DNNs in each system are trained sequentially.
This study does not train through the low-distortion algorithms.

\begin{table*}[t]
\scriptsize
\centering
\sisetup{table-format=2.2,round-mode=places,round-precision=2,table-number-alignment = center,detect-weight=true,detect-inline-weight=math}
\caption{\textsc{One-, Two-, and Six-Microphone SI-SDR, PESQ, eSTOI, PDSAcc, and pSNR Results on The Speech Enhancement Task.}}
\label{resultsspeechenh}
\setlength{\tabcolsep}{3pt}
\begin{tabular}{
clc
S[table-format=2.1,round-precision=1]S[table-format=2.1,round-precision=1]S[table-format=2.1,round-precision=1]
S[table-format=1.1]S[table-format=1.1]!{\enspace}S[table-format=1.1]
S[table-format=2.1,round-precision=1]S[table-format=2.1,round-precision=1]S[table-format=2.1,round-precision=1]
S[table-format=2.1,round-precision=1]S[table-format=2.1,round-precision=1]S[table-format=2.1,round-precision=1]
S[table-format=2.1,round-precision=1]S[table-format=2.1,round-precision=1]S[table-format=2.1,round-precision=1]
}
\toprule
& & \multirow{2}{*}[-6.5pt]{\begin{tabular}[c]{@{}c@{}}Run $\text{DNN}_1$ \\ \#times\end{tabular}} & \multicolumn{3}{c}{SI-SDR (dB)} & \multicolumn{3}{c}{PESQ} & \multicolumn{3}{c}{eSTOI (\%)} & \multicolumn{3}{c}{PDSAcc (\%)} & \multicolumn{3}{c}{pSNR (dB)} \\
\cmidrule(lr{9pt}){4-6} \cmidrule(lr){7-9} \cmidrule(lr){10-12} \cmidrule(lr){13-15} \cmidrule(lr){16-18}
\multicolumn{1}{l}{Entry} & \multicolumn{1}{l}{Systems} & & \multicolumn{1}{c}{1} & \multicolumn{1}{c}{2} & {6} & \multicolumn{1}{c}{1} & \multicolumn{1}{c}{2} & {6} & \multicolumn{1}{c}{1} & \multicolumn{1}{c}{2} & {6} & \multicolumn{1}{c}{1} & \multicolumn{1}{c}{2} & {6} & \multicolumn{1}{c}{1} & \multicolumn{1}{c}{2} & {6} \\

\midrule

0a & Unprocessed & - & -6.2 & {-} & {-} & 1.44 & {-} & {-} & 41.1 & {-} & {-} & {-} & {-} & {-} & -6.6  & {-} & {-} \\

\midrule

1a & SISO$_1$ & 1 & 6.7 & {-} & {-} & 2.67 & {-} & {-} & 80.4 & {-} & {-} & 66.6 & {-} & {-} & 8.6 & {-} & {-} \\
1b & SISO$_1$+SISO$_2$ & 1 & 7.2 & {-} & {-} & 2.75 & {-} & {-} & 82.0 & {-} & {-} & 67.7 & {-} & {-} & 9.2 & {-} & {-} \\
1c & SISO$_1$+WPE+SISO$_3$ & 1 & 8.5 & {-} & {-} & 2.92 & {-} & {-} & 84.6 & {-} & {-} & 70.9 & {-} & {-} & 10.5 & {-} & {-} \\
1d & SISO$_1$+FCP+SISO$_4$ & 1 & 9.0 & {-} & {-} & 2.89 & {-} & {-} & 84.6 & {-} & {-} & \bfseries 71.0 & {-} & {-} & 10.9 & {-} & {-} \\
1e & SISO$_1$+FCP\_WPE+WPE+SISO$_5$ & 1 & \bfseries 9.3 & {-} & {-} & \bfseries 2.96 & {-} & {-} & \bfseries 85.4 & {-} & {-} & 70.9 & {-} & {-} & \bfseries 11.1 & {-} & {-} \\

\hdashline
1f & SISO$_1$+WPE & 1 & -3.8 & {-} & {-} & 1.52 & {-} & {-} & 48.8 & {-} & {-} & 54.6 & {-} & {-} & 2.6 & {-} & {-} \\
1g & SISO$_1$+FCP & 1 & -4.4 & {-} & {-} & 1.46 & {-} & {-} & 44.7 & {-} & {-} & 60.9 & {-} & {-} & 6.6 & {-} & {-} \\
1h & SISO$_1$+FCP\_WPE & 1 & -1.0 & {-} & {-} & 1.49 & {-} & {-} & 49.7 & {-} & {-} & 61.0 & {-} & {-} & 6.6 & {-} & {-} \\

\midrule

2a & MISO$_1$ & 1 & {-} & 7.9 & 9.7 & {-} & 2.90 & 3.09 & {-} & 84.1 & 87.6 & {-} & 69.2 & 71.3 & {-} & 9.7 & 11.4 \\
2b & MISO$_1$+MISO$_3$ & 1 & {-} & 8.1 & 10.2 & {-} & 2.93 & 3.18 & {-} & 84.6 & 88.4 & {-} & 69.5 & 72.9 & {-} & 9.9 & 11.9 \\
2c & MISO$_1$+MVDR+MISO$_2$ & $P$ & {-} & 8.8 & 12.2 & {-} & 3.01 & 3.42 & {-} & 85.9 & 91.0 & {-} & 71.3 & 78.8 & {-} & 10.7 & 14.2 \\
2d & MISO$_1$+MVDR+WPE+MISO$_4$ & $P$ & {-} & 10.5 & 13.5 & {-} & 3.11 & 3.49 & {-} & 87.4 & 91.9 & {-} & 74.0 & 80.8 & {-} & 12.4 & 15.6 \\
\hdashline
2e & MISO$_1$+MVDR & $P$ & {-} & -1.8 & 4.9 & {-} & 1.54 & 1.85 & {-} & 50.2 & 67.2 & {-} & 61.8 & 72.3 & {-} & 4.9 & 9.4 \\
2f & MISO$_1$+WPE & $P$ & {-} & -2.1 & -0.7 & {-} & 1.62 & 1.69 & {-} & 55.4 & 59.8 & {-} & 58.0 & 61.0 & {-} & 3.3 & 4.2 \\
\midrule
3a & MIMO & 1 & {-} & 7.5 & 8.8 & {-} & 2.85 & 3.09 & {-} & 83.6 & 87.1 & {-} & 68.5 & 68.6 & {-} & 9.3 & 10.2 \\
3b & MIMO+MVDR+WPE+MISO$_5$ & 1 & {-} & 10.8 & 13.2 & {-} & 3.12 & 3.48 & {-} & 87.5 & 91.8 & {-} & 74.7 & 80.4 & {-} & 12.8 & 15.3 \\
\hdashline
3c & MIMO+MVDR & 1 & {-} & -2.4 & 3.1 & {-} & 1.54 & 1.85 & {-} & 50.2 & 67.6 & {-} & 61.9 & 71.5 & {-} & 4.9 & 8.9 \\
3d & MIMO+WPE & 1 & {-} & -2.1 & -0.6 & {-} & 1.62 & 1.68 & {-} & 55.4 & 59.8 & {-} & 58.0 & 61.0 & {-} & 3.4 & 4.2 \\
\midrule
4a & MISO$_1$+mMVDR+WPE+MISO$_6$ & 1 & {-} & 10.7 & 13.5 & {-} & 3.17 & 3.53 & {-} & 87.9 & 92.2 & {-} & 74.5 & 81.0 & {-} & 12.6 & 15.6 \\
4b & MISO$_1$+mMVDR\_WPE+WPE+MISO$_7$ & 1 & {-} & 11.1 & 14.6 & {-} & 3.22 & 3.65 & {-} & 88.6 & 93.6 & {-} & 75.6 & 83.7 & {-} & 13.0 & 16.8 \\
4c & MISO$_1$+mWMPDR\_WPE+WPE+MISO$_8$ & 1 & {-} & 11.3 & 15.1 & {-} & 3.24 & 3.71 & {-} & 88.9 & 94.2 & {-} & 76.1 & 84.8 & {-} & 13.3 & 17.3 \\
4d & MISO$_1$+MCWF\_WPE+WPE+MISO$_9$ & 1 & {-} & 11.1 & 15.1 & {-} & 3.28 & 3.72 & {-} & 89.3 & 94.2 & {-} & 75.9 & 84.3 & {-} & 13.1 & 17.3 \\
\hdashline
4f & MISO$_1$+WPE & 1 & {-} & -2.1 & -0.6 & {-} & 1.62 & 1.68 & {-} & 55.4 & 59.8 & {-} & 58.0 & 61.0 & {-} & 3.3 & 4.2 \\
4g & MISO$_1$+mMVDR & 1 & {-} & -4.1 & -1.2 & {-} & 1.57 & 1.90 & {-} & 50.3 & 65.3 & {-} & 59.1 & 66.8 & {-} & 3.4 & 4.7 \\
4h & MISO$_1$+mMVDR\_WPE & 1 & {-} & -0.1 & 3.1 & {-} & 1.79 & 2.30 & {-} & 62.7 & 77.7 & {-} & 62.6 & 71.1 & {-} & 4.1 & 5.9 \\
4i & MISO$_1$+mWMPDR\_WPE & 1 & {-} & -0.3 & 3.1 & {-} & 1.81 & 2.46 & {-} & 63.5 & 80.7 & {-} & 62.8 & 72.1 & {-} & 3.9 & 6.0 \\
4j & MISO$_1$+MCWF\_WPE & 1 & {-} & 5.0 & 11.6 & {-} & 1.88 & 2.43 & {-} & 66.9 & 84.9 & {-} & 68.3 & 79.0 & {-} & 8.0 & 13.9 \\

\midrule

5a & \begin{tabular}{@{}l@{}}MISO$_1$+FCP\_mWMPDR\_WPE+ \\ \,\,\,\,\,\,\,\,\,\,\,\,mWMPDR\_WPE+WPE+MISO$_{10}$\end{tabular} & 1 & {-} & \bfseries 12.0 & \bfseries 15.4 & {-} & \bfseries 3.33 & \bfseries 3.74 & {-} & \bfseries 90.0 & \bfseries 94.4 & {-} & \bfseries 77.1 & \bfseries 85.0 & {-} & \bfseries 13.9 & \bfseries 17.6 \\

\hdashline

5b & MISO$_1$+FCP\_mWMPDR\_WPE & 1 & {-} & 3.4 & 8.2 & {-} & 1.77 & 2.39 & {-} & 65.3 & 82.5 & {-} & 67.8 & 76.7 & {-} & 8.7 & 12.0 \\

\midrule

6a & MISO$_1$+GEV+MISO$_{2a}$ & $P$ & {-} & 8.7 & 11.9 & {-} & 3.00 & 3.36 & {-} & 85.8 & 90.4 & {-} & 71.0 & 77.8 & {-} & 10.6 & 13.9 \\
6b & MISO$_1$+MCWF+MISO$_{2b}$ & $P$ & {-} & 9.6 & 12.7 & {-} & 3.09 & 3.52 & {-} & 86.9 & 92.0 & {-} & 73.2 & 80.1 & {-} & 11.6 & 14.7 \\
\hdashline
6c & MISO$_1$+GEV & $P$ & {-} & -13.2 & -8.5 & {-} & 1.53 & 1.85 & {-} & 47.5 & 63.4 & {-} & 47.4 & 53.3 & {-} & -3.6 & -4.4 \\
6d & MISO$_1$+MCWF & $P$ & {-} & 1.5 & 7.4 & {-} & 1.66 & 2.00 & {-} & 54.5 & 72.9 & {-} & 65.0 & 74.5 & {-} & 6.5 & 11.2 \\

\midrule

7a & Oracle spectral magnitude mask & - & 1.0 & {-} & {-} & 3.31 & {-} & {-} & 90.3 & {-} & {-} & {-} & {-} & {-} & {-} & {-} & {-} \\
7b & Oracle phase-sensitive mask & - & 5.9 & {-} & {-} & 3.55 & {-} & {-} & 90.3 & {-} & {-} & {-} & {-} & {-} & {-} & {-} & {-} \\

\bottomrule
\end{tabular}\vspace{-0.4cm}
\end{table*}

\subsection{Evaluation Metrics}

Our major evaluation metrics include scale-invariant signal-to-distortion ratio (SI-SDR) \cite{LeRoux2019}, extended short-time objective intelligibility (eSTOI) \cite{H.Taal2011}, and perceptual evaluation of speech quality (PESQ).
SI-SDR measures the quality of time-domain sample-level predictions, penalizing a lot if the estimated phase is not correct.
For PESQ, the \textit{python-pesq} toolkit
is used to compute narrow-band MOS-LQO scores based on the ITU P.862.1 standard.
We always use the direct-path signal as the reference for metric computation.
It is obtained by setting the T60 parameter to zero in the RIR generator.
We report word error rates (WER) for ASR.

To measure how accurate the phase-difference signs of the predicted target speech are, we compute their accuracy (denoted as PDSAcc) as follows:
\begin{align}\label{pdsAcc}
\small
\text{PDSAcc} = \frac{ \| \mathcal{I}\Big( \text{Sign}(\angle e^{j(\angle \hat{S}_q - \angle Y_q)}), \text{Sign}(\angle e^{j(\angle S_q - \angle Y_q)}) \Big)\,E_q \|_1}{\|E_q\|_1},
\end{align}
where $\text{Sign}(\cdot)$ operates elementwise and returns $1$ if the value is non-negative and $-1$ otherwise, $\mathcal{I}(\cdot, \cdot)$ operates elementwise and returns $1$ if the two values are equal and $0$ otherwise.
$E_q$ is a binary mask denoting the T-F units with active target speech:
\begin{equation}\label{energymask}
E_q(t,f) = 
\begin{cases}
1,& \text{\,\,\,if\,\,\,} 10\,\text{log}_{10}\,\Big(\frac{|S_q(t,f)|^2}{\text{max}(|S_q|^2)}\Big) \geq -60; \\
0,& \text{\,\,\,otherwise}.
\end{cases}
\end{equation}
Note that due to the randomness of phase-difference signs, a random guess would produce an accuracy of around 50\%.

To measure the performance of phase estimation, we report phase SNR (pSNR) \cite{Wang2021compensation}:
\begin{align}\label{phaseSNR}
\small
\text{pSNR}=10\,\text{log}_{10}\frac{\sum_{t,f} |S_q(t,f)|^2}{\sum_{t,f} \big| S_q(t,f) - |S_q(t,f)|e^{j\angle \hat{S}_q(t,f)} \big|^2},
\end{align}
where oracle magnitude is supplied to emphasize higher-energy T-F units.
We compute it by using $\angle \hat{S}_q$, rather than by using the re-synthesized time-domain signal.
This avoids the influence of the overlap-add algorithm.
pSNR is equal to
\begin{align}
\small
\text{pSNR} \!=\! 10\,\text{log}_{10}\frac{\sum_{t,f} |S_q(t,f)|^2}{\sum_{t,f} 2|S_q(t,f)|^2 \big(1 \!-\! \cos(\angle \hat{S}_q(t,f) \!-\! \angle S_q(t,f))\big)}, \nonumber
\end{align}
essentially measuring how close $\angle \hat{S}_q$ is to $\angle S_q$.

\subsection{Benchmark Systems}

We consider MISO$_1$, MISO$_1$+MISO$_3$, and MISO$_1$+MVDR +MISO$_2$ (see Figs.~\ref{MISO1+MVDR+MISO2} and \ref{MISO1+MISO3}) as the benchmark multi-channel systems, and SISO$_1$ and SISO$_1$+SISO$_2$ (see Fig.~\ref{single_channel_models}(a)) as the baseline monaural systems.
The recent SISO$_1$ and MISO$_1$ networks are strong end-to-end models for single- and multi-channel speech separation \cite{Wang2020d, Wang2020c}.
They model magnitude and phase simultaneously, and share similarities with many contemporary models \cite{Tan2020CSM,Gu2020,ZhangJISI2020,Isik2020,Fu2021,Tan2021,Ren2021,Tzirakis2021}.
Multi-channel systems that use an MVDR beamformer in between two DNNs, such as MISO$_1$+MVDR+MISO$_2$, have shown clearly better performance than using a single end-to-end DNN \cite{Wang2020a, Wang2020b, Wang2020c, Wang2020d}.

\section{Evaluation Results}\label{results}

\subsection{Speech Dereverberation and Enhancement}

Tables~\ref{resultsspeechdereverb} and \ref{resultsspeechenh} respectively report the results on the speech dereverberation and enhancement tasks, along with oracle real-valued T-F masking results based on the spectral magnitude mask ($|S_q|/|Y_q|$) \cite{WYXtrainingtargets} and phase-sensitive mask ($|S_q|/|Y_q|\cos(\angle S_q-\angle Y_q)$) \cite{Erdogan2015}.
The same experiments are done for each task.
We divide the results into multiple blocks separated by solid lines, and index them using different entry numbers in the first column.
In each block, a dashed line is used to separate the results obtained by directly using DNN outputs, and by using the output of the first DNN to drive the intermediate low-distortion algorithms.
This subsection first goes over the SI-SDR, PESQ, eSTOI, PDSAcc, and pSNR results, and then discusses ASR scores in the last paragraph.

Comparing entries 1a, 1b, and 1c, we observe that SISO$_1$+SISO$_2$ only shows slightly better performance than SISO$_1$, and adding a monaural WPE result in SISO$_1$+WPE+\\SISO$_3$ produces clearly better results than SISO$_1$+SISO$_2$, even though the performance of SISO$_1$+WPE in entry 1f is very weak.
A similar trend is observed among MISO$_1$ (2a), MISO$_1$+MISO$_3$ (2b), MISO$_1$+MVDR+MISO$_2$ (2c), and MISO$_1$+MVDR (2e), where an MVDR result is considered as the low-distortion estimate.
Adding a multi-channel  WPE result, MISO$_1$+MVDR+WPE+MISO$_4$ (2d) shows clearly better performance than MISO$_1$+MVDR+MISO$_2$ (2c).
Note that in MISO$_1$+MVDR+WPE+MISO$_4$, the first network has to run $P$ times, once for each microphone.
To avoid this, we proposed MIMO+MVDR+WPE+MISO$_5$ (3b) and MISO$_1$+mMVDR+WPE+MISO$_6$ (4a).
Although MIMO (3a) shows worse performance than MISO$_1$ (2a), MIMO+MVDR+\\WPE+MISO$_5$ (3b) and MISO$_1$+mMVDR+WPE+MISO$_6$ (4a) produce similar performance after using the second DNN.
Using much less computation, both of them show very similar performance to MISO$_1$+MVDR+WPE+MISO$_4$ (2d).
By applying mMVDR beamforming to the WPE results, MISO$_1$+mMVDR\_WPE+WPE+MISO$_7$ in 4b produces better performance than MISO$_1$+mMVDR+WPE+MISO$_6$ in 4a.
We then apply other beamformers including mWMPDR and MCWF to the WPE result.
MISO$_1$+mWMPDR\_WPE+WPE+\\MISO$_8$ in 4c shows better performance on the dereverberation task, and slightly worse performance in some metrics on the enhancement task, compared with MISO$_1$+MCWF\_WPE+WPE+MISO$_9$ in 4d.
With a monaural FCP filter, SISO$_1$+FCP\_WPE+WPE+SISO$_5$ in 1e shows better performance than SISO$_1$+WPE+SISO$_3$ in 1c (and SISO$_1$+FCP+SISO$_4$ in 1d), and MISO$_1$+ FCP\_mWMPDR\_WPE+mWMPDR\_WPE+WPE+MISO$_{10}$~in 5a produces better performance than MISO$_1$+mWM-PDR\_WPE+WPE+MISO$_8$ in 4c.

Our best model in 5a shows large improvement over the MISO$_1$+MVDR+MISO$_2$ baseline in 2c, indicating the effectiveness of including other low-distortion estimates for the second network, and it produces large improvement over the mixture, confirming the effectiveness of our overall approach.

Comparing 4i and 4j of Tables~\ref{resultsspeechdereverb} and \ref{resultsspeechenh}, we notice that MCWF\_WPE shows much better scores than mWMPDRF\_WPE in all the enhancement metrics except PESQ, but the PESQ scores are still competitive.
However, compared with MISO$_1$+mWMPDR\_WPE+WPE+MISO$_8$ in 4c, MISO$_1$+MCWF\_WPE+WPE+MISO$_9$ in 4d produces noticeably worse performance in Table~\ref{resultsspeechdereverb}, and similar performance in Table~\ref{resultsspeechenh}.
This is possibly because the MCWF obtained via Eq.~(\ref{mcwfonwpe}) fits $\hat{S}_q^{(1)}$ well, but does not provide more complementary information than mWMPDR to the second network.

In Table~\ref{resultsspeechdereverb}, we observe that the ASR performance is not strictly correlated with the other enhancement metrics.
This is possibly due to the mismatch between training and testing, as the dereverberation models are trained on our own simulated data but tested on the real-recorded data of REVERB.
One thing we can conclude, though, is that the models leveraging low-distortion estimates produce better performance than the models not leveraging them.

\subsection{Results of Using MVDR, GEV, and MCWF Beamformers}

We investigate the impact of using target estimates with more distortion on the performance of the second DNN.
We use the MISO$_1$+MVDR+MISO$_2$ system in Fig.~\ref{MISO1+MVDR+MISO2}, but replace the MVDR beamformer with a MCWF beamformer or a generalized eigenvector (GEV) beamformer \cite{Warsitz2007}, both of which could generate some speech distortion.
These two systems are denoted as MISO$_1$+MCWF+MISO$_{2a}$ and MISO$_1$+GEV+MISO$_{2b}$.
The MCWF beamforming result is computed by using (\ref{mcwfonwpe}) and (\ref{mcwfonwperesult}), but we replace $\hat{\mathbf{S}}^{\text{WPE}}(t,f)$ with $\mathbf{Y}(t,f)$.
The GEV beamformer is computed as
\begin{align}\label{gevfilter}
\hat{\mathbf{w}}(f)=\mathcal{P}\Big(
\Big(\hat{\mathbf{\Phi}}^{(v)}(f)\Big)^{-1}
\hat{\mathbf{\Phi}}^{(s)}(f)
\Big),
\end{align}
where $\hat{\mathbf{\Phi}}^{(s)}(f)$ and $\hat{\mathbf{\Phi}}^{(v)}(f)$ are computed using Eqs.~(\ref{covariancematrixS}) and (\ref{covariancematrixV}).
A post-filter \cite{Warsitz2007} is used to reduce the distortion on magnitude
\begin{align}\label{ban}
\hat{c}(f)=\frac{\sqrt{\hat{\mathbf{w}}(f)^{\H} \hat{\mathbf{\Phi}}^{(v)}(f) \hat{\mathbf{\Phi}}^{(v)}(f) \hat{\mathbf{w}}(f) /P}}{\hat{\mathbf{w}}(f)^{\H} \hat{\mathbf{\Phi}}^{(v)}(f) \hat{\mathbf{w}}(f)}.
\end{align}
The beamforming result is obtained as
\begin{align}\label{gevresult}
\hat{S}_q^{\text{GEV}}(t,f) = \hat{c}(f) \hat{\mathbf{w}}(f)^{\H} \mathbf{Y}(t,f).
\end{align}

MVDR and MCWF beamformers essentially use the same projection direction \cite{Gannot2017}.
That is, the oracle beamforming filters, with microphone $q$ as the reference microphone, can be written as $\mathbf{w}(f;q) = z(f;q)\mathbf{\Phi}^{(v)}(f)^{-1} \mathbf{d}(f)/d_q(f)$, where $z(f;q)$ is a real-valued spectral gain.
For oracle MVDR, $z(f;q)$ is computed such that $\mathbf{w}(f;q)^{\H} \mathbf{S}(t,f)$ is equal to $S_q(t,f)$, resulting in a distortionless response.
For oracle MCWF, the phase of $\mathbf{w}(f;q)^{\H} \mathbf{S}(t,f)$ is equal to that of $S_q(t,f)$, but a slight distortion is introduced to the target magnitude in order to better suppress non-target signals \cite{Gannot2017}.
For GEV, $z(f;q)$ is a complex value and would introduce a random phase shift at each frequency \cite{Warsitz2007}.
The real-valued post-filter only reduces the distortion on target magnitude, but not on phase.
GEV would therefore introduce more distortion to target speech than MVDR and MCWF.
This can be observed from its low SI-SDR and pSNR scores by comparing entry 6c with 2e and 6d.
From 2c, 6a, and 6b, we observe that MISO$_1$+GEV+MISO$_{2a}$ shows worse performance than the other two, indicating the benefits of using target estimates with lower distortion.
However, the performance gap is not drastic, %
possibly because the distortion introduced by GEV is a time-invariant one-tap complex gain per frequency, and the DNN could implicitly cancel  out the gain through its instance normalization layers.
On the other hand, MISO$_1$+MCWF+MISO$_{2b}$ obtains better performance than MISO$_1$+MVDR+MISO$_2$.
This is possibly because, ideally, MCWF does not distort target phase, similarly to MVDR, and the estimated MCWF obtained via (\ref{mcwfonwpe}) can better suppress non-target signals than the estimated MVDR beamformer.

\subsection{Improvement of Phase-Difference Signs}

In both tasks, we observe clear phase improvement by leveraging low-distortion estimates.
For example, in Table~\ref{resultsspeechdereverb}, MISO$_1$+mWMPDR\_WPE+WPE+MISO$_8$ in 4c produces clearly better PDSAcc and pSNR than MISO$_1$+MISO$_3$ in 2b (88.2\% vs.\ 77.3\% PDSAcc, and 20.1 vs.\ 13.5 dB pSNR), while MISO$_1$+MISO$_3$ is only slightly better than MISO$_1$ (77.3\% vs.\ 76.2\% PDSAcc, and 13.5 vs.\ 12.9 dB pSNR).

\begin{figure}
  \centering  
  \includegraphics[width=9cm]{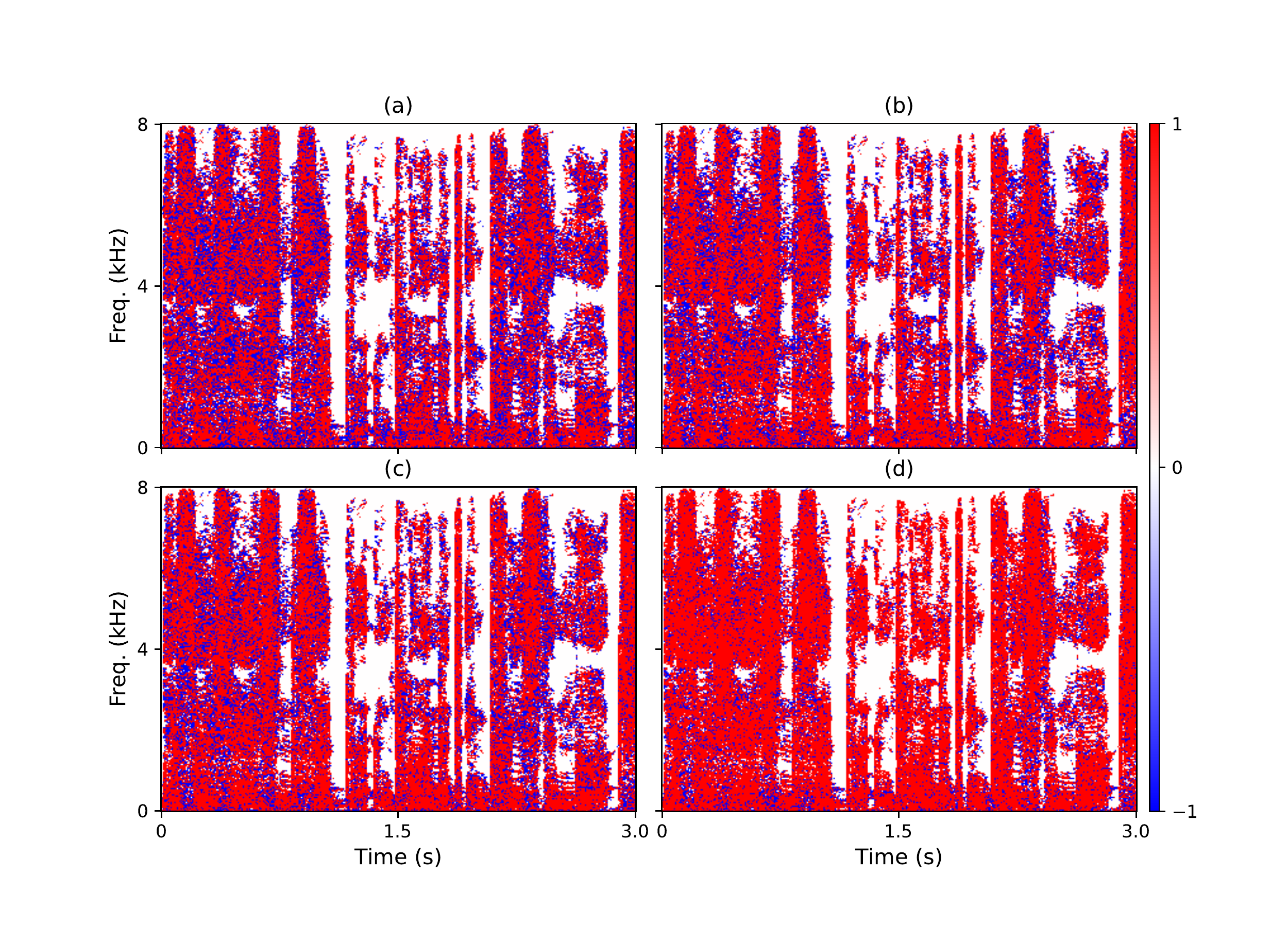}\vspace{-0.2cm}
  \caption{
  Illustration of accuracy of phase-difference signs of estimated target speech of (a) SISO$_1$; (b) MISO$_1$; (c) MISO$_1$+MISO$_3$; and (d) MISO$_1$+ MCWF\_WPE+WPE+MISO$_9$ models. 
  Best viewed in color.
  In this example, the mixture SI-SDR
is -4.9 dB, and the models respectively obtain 6.8, 10.5, 10.7, and 15.9 dB SI-SDR, and 65.6\%, 72.2\%, 72.9\%, and 84.9\% PDSAcc.
  }
  \label{pdscomparison}\vspace{-0.6cm}
\end{figure}

Fig.~\ref{pdscomparison} plots the accuracy of the phase-difference signs of the predicted target speech by various models on a portion of the example mixture used in Fig.~\ref{patternsfigure}, which was sampled from the test set of the enhancement task.
The figures plot $\text{Sign}(\angle e^{j(\angle \hat{S}_q - \angle Y_q)})\text{Sign}(\angle e^{j(\angle S_q - \angle Y_q)})E_q$ (see Eqs.~(\ref{pdsAcc}) and (\ref{energymask}) for the definitions), which is $1$ at a T-F unit if the sign of the predicted speech is correct, $-1$ if wrong, and $0$ if the T-F unit does not contain active target speech.
We notice that PDSAcc is clearly higher in Fig.~\ref{pdscomparison}(d) than in other figures (i.e., more red and fewer blue T-F units), especially around the boundary of each phoneme, where the input SNR is lower.
Although, as illustrated in Fig.~\ref{patternsfigure}(g), the true phase-difference signs are extremely random, Fig.~\ref{pdscomparison}(d) shows that our algorithms can make reasonably good predictions.
These results indicate the effectiveness of exploiting low-distortion target estimates at improving phase-difference signs.

\section{Conclusion}\label{conclusion}

We have proposed a 2stage-DNN system with a linear, low-distortion module in between.
The low-distortion results are computed based on the outputs of the first network, and used as extra features to the second network.
Evaluation results on simulated speech dereverberation and enhancement tasks show that all the considered low-distortion algorithms can 
to various degrees improve the second network.
Among them, the convolutional beamformer with additional FCP filtering %
leads to the largest improvement on the dereverberation task, and produces very strong performance on the enhancement task.
Compared to the previous MISO-BF-MISO system \cite{Wang2020c, Wang2020d}, the proposed systems avoid the reliance on uniform circular array geometry as well as unnecessary computations in the first network, while obtaining clearly better performance.

End-to-end DNNs can dramatically improve the performance of speech enhancement, and their immediate outputs typically exhibit better enhancement scores than the outputs of linear-filtering algorithms such as WPE, FCP, and time-invariant beamforming.
Our study shows that the linear-filtering results, when used as extra features, can dramatically improve the DNN models.
We have given a novel explanation on the benefits of leveraging such low-distortion algorithms.
That is, the low-distortion algorithms can produce a reliable, non-aggressive enhancement result that is closer to the target speech than the mixture, and such a reliable result, when used as extra features to a DNN, can help the DNN better predict the target magnitude and phase (including the absolute phase difference and phase-difference sign).
This understanding, we believe, is very valuable, as it reveals the strong potential of combining DNNs with conventional low-distortion algorithms, and points out a promising direction for future research.

Only WPE, FCP, and beamforming are explored here for low-distortion target estimation, considering their popularity. 
Our conjecture is that the results of many low-distortion enhancement algorithms could 
to some extent improve the second DNN.
Future research shall consider other low-distortion algorithms such as \cite{Nakatani2021, Boeddeker2021}, and design novel end-to-end DNN architectures in light of this novel understanding.

\section*{Acknowledgment}\label{acknowledgment}

We would like to thank Dr. Shinji Watanabe for helpful discussions.

\bibliographystyle{IEEEtran}
\bibliography{references.bib}

\end{document}